\documentclass{article}
\usepackage[english]{babel}
\usepackage[a4paper,top=2cm,bottom=2cm,left=3cm,right=3cm,marginparwidth=1.75cm]{geometry}
\usepackage{amsmath}
\allowdisplaybreaks
\usepackage{mathtools}
\usepackage{xcolor}             
\usepackage{amsthm}             
\usepackage{amsfonts}           
\usepackage{tcolorbox}          
\usepackage{graphicx}
\usepackage{physics}
\usepackage[colorlinks=true, allcolors=blue]{hyperref}
\usepackage{amssymb}
\usepackage{authblk}

\usepackage[nocompress]{cite}
\usepackage{cleveref}           

\theoremstyle{remark}

\theoremstyle{theorem}
\newtheorem{theorem}{Theorem}[section]
\newtheorem{proposition}{Proposition}[section]

\newtheorem{lemma}[theorem]{Lemma}

\DeclareMathOperator*{\res}{res}
\DeclareMathOperator{\erm}{e} 
\DeclareMathOperator{\irm}{i} 
\DeclareMathOperator{\Ham}{\mathcal{H}}
\DeclareMathOperator{\bbl}{\mathbf{l}}
\DeclareMathOperator{\bbi}{\mathbf{i}}

\title{Fermionic Basis in Conformal Field Theory: The Free Fermion Point}
\author[1]{S. Adler}
\author[1]{H. Boos}
\affil[1]{Fakult\"{a}t f\"{u}r Mathematik und Naturwissenschaften, \protect\\ Bergische Universit\"{a}t Wuppertal, 42097 Wuppertal, Germany}
\date{}

\begin{document}
\maketitle

\begin{abstract}
In this work, we use the master function approach developed in \cite{BG_2012} to describe the CFT limit of the six-vertex model at the free fermion point. Using the ODE/IM correspondence, we obtain an explicit form of the master function. This allows us to compute the asymptotic expansion of the function $\omega(\lambda, \mu)$ describing the expectation values of the fermionic basis operators \cite{HGS_1, HGS_2, HGS_3}. As a result, we describe the entire Virasoro module of the corresponding CFT, including the integrals of motion. 
\end{abstract}

\section{Introduction}
\label{sec:introduction}
In papers \cite{HGS_4, Boos_TBAE, AB_2025} some connections between the CFT with the central charge $c=1-6\nu^2/(1-\nu)$  and the XXZ model with the deformation parameter $q=\erm^{\irm \pi \nu}$ were established. 
The correlation functions of the corresponding six-vertex model on a cylinder were described in terms of certain fermionic operators and two functions $\rho$ and $\omega$. The CFT can be obtained in the scaling limit, in which these functions become $\rho^{\text{sc}}(\lambda|\kappa,\kappa')$, $\omega^{\text{sc}}(\lambda,\mu|\alpha,\kappa,\kappa')$ depending on the spectral parameters $\lambda$ and $\mu$, ``disorder field'' $\alpha$ and two ``flux'' parameters $\kappa$ and $\kappa'$ taken on the left and right boundaries of the cylinder. To obtain the correlation functions one needed to analyse the asymptotic expansions of the above functions at large spectral parameters $\lambda,\mu \to \infty$. To this end, the Wiener-Hopf factorisation technique was applied. Unfortunately, this could be carried out only modulo integrals of motion, effectively restricting one to the case $\kappa=\kappa'$. Some attempts to escape this condition and take into account the contribution of integrals of motion were done in \cite{BS_2016}. On a technical level, this required to solve certain functional equations. While it was possible to do for several levels of the Virasoro module, the general case remains highly challenging.  
	
In this paper, we show how the action of the integrals of motion can be incorporated in the final answer in the case of the free fermionic model, corresponding to $\nu=1/2$.

Let us give a short overview of the established relation between the CFT and six-vertex model. It was found to be valid that:
\begin{align}\label{eq:introduction-CFT-6v}
\frac{\langle \Delta_{-} | P_\alpha(\mathbf{l}_{-k})\phi_\alpha(0) |\Delta_{+} \rangle}
	 {\langle \Delta_{-} | \phi_\alpha(0) |\Delta_{+} \rangle}
= 
\lim\limits_{\mathfrak{n}\rightarrow\infty, l \rightarrow 0, \mathbf{n} l = 2\pi R}
	Z^{\kappa, s}\left\{ q^{2\alpha S(0)} \mathcal{O} \right\}.
\end{align}
The left-hand side of \eqref{eq:introduction-CFT-6v} contains the CFT data. It is a normalised three-point function of the CFT defined on a cylinder of radius $R$ parametrised by a complex variable $z=x+\irm y$ with spatial coordinate $x: -\infty < x < +\infty$ and the coordinate in the Matsubara direction $y: -\pi R < y <\pi R$. The periodic boundary conditions, that is, the equivalence of the points $x\pm\irm\pi R$, are implied. At the origin $z=0$ some descendant of a primary field $\phi_\alpha(0)$ with conformal dimension
\begin{align}
\Delta_\alpha = \frac{\nu^2 \alpha(\alpha-2)}{4(1-\nu)}
\end{align}
is inserted. The polynomial $P_\alpha(\mathbf{l}_{-k})$ describes the corresponding combination of the Virasoro algebra generators $\mathbf{l}_{-k}$, $k>0$. We call these Virasoro generators ``local'' in a sense that they are defined in the vicinity of $z=0$ with the corresponding energy-momentum tensor
\begin{align}
T(z) = \sum_{n=-\infty}^{+\infty} \mathbf{l}_n z^{-n-2}.
\end{align}
The left and right states $|\Delta_+ \rangle$ and $\langle \Delta_- |$ correspond to two primary fields $\phi_{\pm}$ with conformal dimensions $\Delta_{\pm}$, inserted at $x=\pm\infty$. They are defined in such a way that $L_n|\Delta_+ \rangle = \delta_{n, 0}\Delta_+ |\Delta_+ \rangle$, $n\geq0$ when $x=\infty$ and $\langle \Delta_- |L_n= \delta_{n, 0}\Delta_- \langle \Delta_- |$, $n\leq0$ at $x=-\infty$. We call the Virasoro generators $L_n$ ``global''. In \cite{Zamolodchikov_1987} local integrals of motion were introduced. They act on local operators as
\begin{align}
(\mathbf{i}_{2n-1}\mathcal{O})(z)=\int\limits_{\mathcal{C}_z}\frac{dw}{2\pi\mathrm{i}}h_{2n}(w)\mathcal{O}(z),
\end{align}
where the densities $h_{2n}(w)$ are certain descendants of the identity operator. An important property is that
\begin{align}
\langle\Delta_{-}| (\mathbf{i}_{2n-1}\mathcal{O})(z) |\Delta_{+}\rangle = (I_{2n-1}^{+} - I_{2n-1}^{-}) \langle\Delta_{-}| \mathcal{O}(z) |\Delta_{+}\rangle.
\end{align}
where $I^{\pm}_{2n-1}$ are vacuum eigenvalues of the local integrals of motion on the Verma module with conformal dimensions $\Delta_{\pm}$. The whole Verma module is spanned by the elements
\begin{align*}
	\mathbf{i}_{2k_1-1}\ldots\mathbf{i}_{2k_p-1} \mathbf{l}_{-2m_1}\ldots\mathbf{l}_{-2m_q}(\phi_{\alpha}(0)).
\end{align*}

The right-hand side of \eqref{eq:introduction-CFT-6v} contains the lattice model data. In order to describe it, we need to introduce the fermionic basis constructed in \cite{HGS_1, HGS_2, HGS_3}. It consists of the creation operators $\mathbf{t}^{*}, \mathbf{b}^{*}, \mathbf{c}^{*}$ together with the annihilation operators $\mathbf{b}, \mathbf{{c}}$ that act on the space
\begin{align}
	\mathcal{W}^{(\alpha)}=\bigoplus\limits_{s=-\infty}^{\infty} \mathcal{W}_{\alpha-s, s},
\end{align}
where $\mathcal{W}_{\alpha-s, s}$ is the subspace of quasi-local operators of the spin $s$ with the shifted $\alpha$-parameter. The operators forming the fermionic basis are defined as formal power series of the spectral parameter $\zeta^2$ around $\zeta^2=1$ and have the block structure
\begin{align}
	&\mathbf{t}^{*}(\zeta)                   : \mathcal{W}_{\alpha-s, s}   \rightarrow \mathcal{W}_{\alpha-s, s}, \\
	&\mathbf{b}^{*}(\zeta), \mathbf{c}(\zeta): \mathcal{W}_{\alpha-s+1, s-1} \rightarrow \mathcal{W}_{\alpha-s, s}, \cr
	&\mathbf{c}^{*}(\zeta), \mathbf{b}(\zeta): \mathcal{W}_{\alpha-s-1, s+1} \rightarrow \mathcal{W}_{\alpha-s, s}. \nonumber
\end{align}
The operator $\mathbf{t}^{*}(\zeta)$ plays the role of a generating function of the commuting integrals of motion. In a sense it is bosonic. It commutes with all fermionic operators $\mathbf{b}(\zeta), \mathbf{c}(\zeta)$ and $\mathbf{b}^{*}(\zeta), \mathbf{c}^{*}(\zeta)$ which obey canonical anti-commutation relations
\begin{align}\label{Reminder_3}
	[\mathbf{c}(\xi), \mathbf{c}^{*}(\zeta)]_{+}=\psi(\xi/\zeta, \alpha),
	\quad
	[\mathbf{b}(\xi), \mathbf{b}^{*}(\zeta)]_{+}=-\psi(\zeta/\xi, \alpha).
\end{align}
The rest of the anti-commutation relations are zero. The function $\psi(\zeta, \alpha)$ is given by
\begin{align}\label{psi}
	\psi(\zeta, \alpha)=\frac{\zeta^{\alpha}}{\zeta^2-1}.
\end{align}
The operator $q^{2\alpha S(0)}$ can be treated as a ``primary field'': the annihilation operators act on it as zero
\begin{align}
	\mathbf{b}(\zeta)(q^{2\alpha S(0)})=0, 
	\quad
	\mathbf{c}(\zeta)(q^{2\alpha S(0)})=0,
\end{align}
and the space of states is generated via multiple action of creation operators $\mathbf{t}^{*}, \mathbf{b}^{*}, \mathbf{c}^{*}$. In \cite{Boos_2009} it was proved that, for the homogeneous XXZ model, these operators form a basis of the space of quasi-local operators.

In the right-hand side of \eqref{eq:introduction-CFT-6v} we take the scaling limit of a normalised partition function of the six-vertex model on a cylinder with insertion of a quasi-local operator $q^{2\alpha S(0)}\mathcal{O}$:
\begin{align}\label{Reminder_2}
	Z^{\kappa, s}\left\lbrace q^{2\alpha S(0)} \mathcal{O}\right\rbrace=
	\frac{\tr_\mathrm{S} \tr_\mathbf{M} \left(Y_{\mathbf{M}}^{(-s)} T_{\mathrm{S}, \mathbf{M}} q^{2\kappa S} \mathbf{b}^{*}_{\infty, s-1} \ldots \mathbf{b}^{*}_{\infty, 0}\left(q^{2\alpha S(0)}\mathcal{O}\right) \right)}
	{\tr_\mathrm{S} \tr_\mathbf{M} \left(Y_{\mathbf{M}}^{(-s)} T_{\mathrm{S}, \mathbf{M}} q^{2\kappa S} \mathbf{b}^{*}_{\infty, s-1} \ldots \mathbf{b}^{*}_{\infty, 0}\left(q^{2\alpha S(0)}\right) \right)}.
\end{align}
Above $\mathbf{b}^{*}_{\infty, j}$ are the ``lattice screening operators'', which are the coefficients in the expansion of the singular part of $\mathbf{b}^*$ at $\zeta^2=0$. They increase the spin of $\mathcal{O}$ by $s$. This is compensated by the operator $Y_{\mathbf{M}}^{(-s)}$ at the boundary which ensures that the ice-rule of the six vertex model is satisfied. The monodromy matrix $T_{\mathrm{S}, \mathbf{M}}$ is defined as a tensor product of evaluation representations of $U_q(\widehat{\mathfrak{sl}_2})$:
\begin{align}
	T_{\mathrm{S}, \mathbf{M}} = \overset{\curvearrowright}{\prod_{j=-\infty}^{+\infty}} T_{j, \mathbf{M}}, 
	\quad
	T_{j, \mathbf{M}} \equiv T_{j, \mathbf{M}}(1),
	\quad
	T_{j, \mathbf{M}}(\zeta)=\overset{\curvearrowleft}{\prod_{\mathbf{m}=1}^{\mathbf{n}}} L_{j, \textbf{m}}(\zeta),
\end{align}
where $L$ is the standard $L$-operator of the six vertex model
\begin{align}
	L_{j, \textbf{m}}(\zeta)=q^{-\frac{1}{2}\sigma_{j}^{3}\sigma_{\textbf{m}}^3} - \zeta^{2}q^{\frac{1}{2}\sigma_{j}^3\sigma_\textbf{m}^{3}} - \zeta(q-q^{-1})(\sigma_j^{+}\sigma_\textbf{m}^{-}+\sigma_{j}^{-}\sigma_{\textbf{m}}^{+}).
\end{align}

As it was discussed in \cite{HGS_4}, the functional \eqref{Reminder_2} does not depend on the concrete choice of the screening operator $Y_{\mathbf{M}}^{(-s)}$. In the case of the infinite lattice, one can also change the boundary conditions and, instead of taking the traces in the right-hand side of \eqref{Reminder_2}, insert two one-dimensional projectors $|\kappa\rangle \langle\kappa|$ and $|\kappa+\alpha-s, s\rangle \langle\kappa+\alpha-s, s|$ at the boundary, where the vector $|\kappa\rangle$ is the eigenvector of the twisted transfer matrix $T_{\mathbf{M}}(\zeta, \kappa)$ with the maximal eigenvalue $T(\zeta, \kappa)$ in the zero spin sector. Similarly, the vector $|\kappa+\alpha-s, s\rangle$ is the eigenvector of the twisted transfer matrix $T_{\mathbf{M}}(\zeta, \kappa+\alpha-s)$ with the maximal eigenvalue $T(\zeta, \kappa+\alpha-s, s)$ in the sector with spin $s$. Then
\begin{align}\label{Reminder_4}
	Z^{\kappa, s}\left\lbrace q^{2\alpha S(0)} \mathcal{O} \right\rbrace \rightarrow
	\frac{\langle \kappa+\alpha-s, s| T_{\mathrm{S}, \mathbf{M}} q^{2\kappa S} \mathbf{b}^{*}_{\infty, s-1} \ldots \mathbf{b}^{*}_{\infty, 0}\left(q^{2\alpha S(0)}\mathcal{O}\right) |\kappa\rangle}
	{\langle \kappa+\alpha-s, s| T_{\mathrm{S}, \mathbf{M}} q^{2\kappa S} \mathbf{b}^{*}_{\infty, s-1} \ldots \mathbf{b}^{*}_{\infty, 0}\left(q^{2\alpha S(0)}           \right) |\kappa\rangle}.
\end{align}

The theorem of Jimbo, Miwa and Smirnov \cite{HGS_3} states that
\begin{align}\label{Reminder_JMS}
	&Z^{\kappa, s} \left\lbrace \mathbf{t}^{*}(\zeta)(X) \right\rbrace
	= 2\rho(\zeta; \kappa, \kappa+\alpha, s) Z^{\kappa, s}\lbrace X \rbrace, \\
	&Z^{\kappa, s} \left\lbrace \mathbf{b}^{*}(\zeta)(X) \right\rbrace
	= \frac{1}{2\pi i} \oint_\Gamma \omega(\zeta, \xi; \kappa, \alpha, s) Z^{\kappa, s} \lbrace\mathbf{c}(\xi)(X)\rbrace\frac{d\xi^2}{\xi^2}, \cr
	&Z^{\kappa, s} \left\lbrace \mathbf{c}^{*}(\zeta)(X) \right\rbrace
	= -\frac{1}{2\pi i} \oint_\Gamma \omega(\xi, \zeta; \kappa, \alpha, s) Z^{\kappa, s} \lbrace\mathbf{b}(\xi)(X)\rbrace\frac{d\xi^2}{\xi^2} \nonumber,
\end{align}
where the contour $\Gamma$ goes around all singularities of the integrand except $\xi^2=\zeta^2$. The direct consequence of the above formula and the anti-commutation relations \eqref{Reminder_3} is the determinant formula
\begin{align}\label{Reminder_JMS_Det}
	&Z^{\kappa, s}\lbrace \mathbf{t}^{*}(\zeta^0_1) \ldots \mathbf{t}^{*}(\zeta^0_p)
	\mathbf{b}^{*}(\zeta^+_1) \ldots \mathbf{b}^{*}(\zeta^+_r)
	\mathbf{c}^{*}(\zeta^-_r) \ldots \mathbf{c}^{*}(\zeta^-_1)
	(q^{2\alpha S(0)})
	\rbrace \\
	&=\prod_{i=1}^{p} 2\rho(\zeta^0_i; \kappa, \kappa+\alpha, s) \times \det(\omega(\zeta^+_i, \zeta^-_j; \kappa, \alpha, s))_{i, j=1, \ldots, r}. \nonumber
\end{align}
The functions $\rho$ and $\omega$ are completely defined by the Matsubara data. The function $\rho$ is the ratio of two eigenvalues of the transfer matrix
\begin{align}
	\rho(\zeta; \kappa+\alpha-s, s)=\frac{T(\zeta, \kappa+\alpha-s, s)}{T(\zeta, \kappa)}.
\end{align}
The function $\omega$ will be the main object of our study. Generally speaking, its definition is quite involved, but for this work we will only need its relation \eqref{omega_HHPhi} to the so-called master function which we will introduce in \Cref{sec:master-function}.

The scaling limit taken for the right-hand side of \eqref{eq:introduction-CFT-6v} in the Matsubara direction means
\begin{align}
	\mathbf{n}\rightarrow\infty,
	\quad
	a\rightarrow 0
	\quad
	\mathbf{n}a=2\pi R.
\end{align}
Here $a$ is the step of the lattice and $R$ is the radius of the cylinder which is fixed. Simultaneously one should rescale the spectral parameter
\begin{align}
	\zeta=\lambda\bar{a}^\nu, 
	\quad
	\bar{a}=Ca,
\end{align}
where $C$ is some fine-tuning constant, which is needed to compare the scaling limit to CFT. One of the most important points of \cite{HGS_4} was to define the scaling limits of $\rho$ and $\omega$
\begin{align}
	&\rho^{\text{sc}}(\lambda; \kappa, \kappa')=
	\lim\limits_{\substack{\textbf{n}\rightarrow\infty,\\ a\rightarrow 0,\\ \mathbf{n}a=2\pi R}}
	\rho(\lambda\bar{a}^\nu; \kappa, \alpha, s), \\
	&\omega^{\text{sc}}(\lambda, \mu; \kappa, \kappa', \alpha) = \frac{1}{4}
	\lim\limits_{\substack{\textbf{n}\rightarrow\infty,\\ a\rightarrow 0,\\ \mathbf{n}a=2\pi R}}
	\omega(\lambda\bar{a}^\nu, \mu\bar{a}^\nu; \kappa, \alpha, s),
\end{align}
where $\kappa'$ is defined through an analogue of the Dotsenko-Fateev condition
\begin{align*}
	\kappa'=\kappa+\alpha+2\frac{1-\nu}{\nu}s.
\end{align*}
The continuum limit can be taken in both directions along the cylinder.

It was proposed in \cite{HGS_4} that the creation operators are well-defined in the scaling limit for the space direction when the spatial coordinate $x=ja$ for a site $j$ is finite
\begin{align*}
	\boldsymbol{\tau}^{*}(\lambda)  =\lim\limits_{a\rightarrow 0}\frac{1}{2}\mathbf{t}^* (\lambda \bar{a}^\nu),
	\quad
	\boldsymbol{\beta}^{*}(\lambda) =\lim\limits_{a\rightarrow 0}\frac{1}{2}\mathbf{b}^* (\lambda \bar{a}^\nu),
	\quad
	\boldsymbol{\gamma}^{*}(\lambda)=\lim\limits_{a\rightarrow 0}\frac{1}{2}\mathbf{c}^* (\lambda \bar{a}^\nu),
\end{align*}
and the lattice primary field scales to the CFT primary field
\begin{align*}
	\phi_\alpha(0)\phi_{\alpha}^{-1}(-\infty)=\lim_{a\rightarrow0} q^{2\alpha S(0)}.
\end{align*}
The asymptotic expansions of the above operators at $\lambda\rightarrow\infty$ read
\begin{align}\label{eq:introduction-fb-asymptotics}
\begin{aligned}
	&\log(\boldsymbol{\tau}^{*} (\lambda))                  \sim \sum_{j=1}^{\infty} \boldsymbol{\tau}^{*}_{2j-1} \lambda^{-\frac{2j-1}{\nu}}, \\
	&\frac{1}{\sqrt{\boldsymbol{\tau}^{*}}} \boldsymbol{\beta}^{*}(\lambda) \sim \sum_{j=1}^{\infty} \boldsymbol{\beta}^{*}_{2j-1} \lambda^{-\frac{2j-1}{\nu}}, 
	\quad
	\frac{1}{\sqrt{\boldsymbol{\tau}^{*}}} \boldsymbol{\gamma}^{*}\sim \sum_{j=1}^{\infty} \boldsymbol{\gamma}^{*}_{2j-1} \lambda^{-\frac{2j-1}{\nu}}.
\end{aligned}
\end{align}

For the scaling limit of the functional \eqref{Reminder_JMS_Det} the JMS theorem can also be written in the determinant form
\begin{align}
	&Z_R^{\kappa, \kappa'} \{ 
	\boldsymbol{\tau}^{*}(\lambda^0_1)\ldots \boldsymbol{\tau}^{*}(\lambda^0_p)
	\boldsymbol{\beta}^{*}(\lambda^+_1)\ldots \boldsymbol{\beta}^{*}(\lambda^+_r)
	\boldsymbol{\gamma}^{*}(\lambda^-_r)\ldots \boldsymbol{\gamma}^{*}(\lambda^-_1)
	(\Phi_\alpha(0))
	\}=\\
	&\qquad \prod_{i=1}^{p}\rho^{\text{sc}}(\lambda^0_i; \kappa, \kappa') \times \det(\omega^{\text{sc}}(\lambda^+_i, \lambda^-_j; \kappa, \kappa', \alpha))_{i,j=1, \ldots, r}. \nonumber
\end{align}

\section{The master function and ODE}\label{sec:master-function}
\subsection{Baxter TQ-relation}
Let us begin with the case of a lattice with an even finite number of sites $\mathbf{n}$ in the Matsubara direction. The Baxter TQ-relation for the eigenvalues $T(\lambda, \kappa, s)$ and $Q(\lambda, \kappa, s)$ of the transfer matrix $T_{\mathbf{M}}(\lambda, \kappa, s)$ and the $Q$-operator $Q_{\mathbf{M}}(\lambda, \kappa, s)$ is
\begin{align}\label{eq:TQ-Lattice}
T(\lambda, \kappa, s)Q(\lambda, \kappa, s) = d(\lambda)Q(q\lambda, \kappa, s) + a(\lambda) Q(q^{-1}\lambda, \kappa, s),
\qquad
Q(\lambda, \kappa, s)=\lambda^{-\kappa+s}A(\lambda, \kappa, s).
\end{align}
The functions $T(\lambda, \kappa, s)$ and $A(\lambda, \kappa, s)$ depend polynomially on the spectral parameter $\lambda$ in every spin sector $s$, $0 \leq s \leq \frac{\mathbf{n}}{2}$, and
\begin{align}
a(\lambda) = (1-q\lambda^2)^{\mathbf{n}}, 
\qquad
d(\lambda) = (1-q^{-1}\lambda^2)^{\mathbf{n}}.
\end{align}
The Bethe ansatz equations follow from the TQ-relation. Since the left-hand side of \eqref{eq:TQ-Lattice} vanishes when taken at the Bethe roots, the auxiliary function defined as
\begin{align}\label{eq:Auxiliary-Function}
\mathfrak{a}(\lambda, \kappa, s) = \frac{d(\lambda)Q(q\lambda, \kappa, s)}{a(\lambda) Q(q^{-1}\lambda, \kappa, s)}
\end{align}
satisfies the Bethe ansatz equations
\begin{align}\label{eq:BAE}
\mathfrak{a}(\lambda_r, \kappa, s) = -1,
\qquad
r = 1, \ldots, \frac{\mathbf{n}}{2}-s.
\end{align}

Now we consider the conformal limit of the lattice that was discussed in \Cref{sec:introduction}:
\begin{align}
& T^{\text{sc}}(\lambda, \kappa) = \lim\limits_{\mathfrak{n}\rightarrow\infty, l \rightarrow 0, \mathbf{n} l = 2\pi R}
T(\lambda \bar{l}^\nu, \kappa), \\
&Q^{\text{sc}}(\lambda, \kappa) = \lim\limits_{\mathfrak{n}\rightarrow\infty, l \rightarrow 0, \mathbf{n} l = 2\pi R}
Q(\lambda \bar{l}^\nu, \kappa), 
\end{align} 
where $\bar{l}=l\cdot C(\nu)$, $C(\nu) = \frac{\Gamma\left(\frac{1-\nu}{2\nu}\right)}{2\sqrt{\pi}}\Gamma\left(\frac{1}{2\nu}\right)\Gamma(\nu)^{\frac{1}{\nu}}$. In the current work we consider the free fermionic point corresponding to the value $\nu=\frac{1}{2}$. This means, in particular, that the functions $a(\lambda)$ and $d(\lambda)$ have the following scaling limit
\begin{align}
& a(\lambda \bar{l}^\nu) = (1 - q \bar{l}\lambda^2) = \left(1 - q \frac{2\pi R C}{\mathbf{n}} \lambda^2 \right)^\mathbf{n} \xrightarrow[\text{scaling}]{} \erm^{-\irm\pi^2\lambda^2} \\
& d(\lambda \bar{l}^\nu) = (1 - q^{-1} \bar{l}\lambda^2) = \left(1 - q^{-1} \frac{2\pi R C}{\mathbf{n}} \lambda^2 \right)^\mathbf{n} \xrightarrow[\text{scaling}]{} \erm^{\irm\pi^2\lambda^2}.
\end{align}
The TQ-relation \eqref{eq:TQ-Lattice} in the scaling limit takes the form
\begin{align}
T^{sc}(\lambda, \kappa)Q^{sc}(\lambda, \kappa, s) = \erm^{\irm \pi^2 \lambda^2}Q^{sc}(q\lambda, \kappa, s) + \erm^{-\irm \pi^2 \lambda^2} Q^{sc}(q^{-1}\lambda, \kappa, s).
\end{align}
Finally, due to cancellation of the functions $A(\lambda, \kappa)$ in \eqref{eq:Auxiliary-Function}, the scaling limit of the auxiliary function $\mathfrak{a}(\lambda, \kappa)$ is simply
\begin{align}
\mathfrak{a}^{\text{sc}}(\lambda, \kappa)=\erm^{-\irm\pi\kappa+2\irm\pi^2\lambda^2}.
\end{align}
This, together with the equation \eqref{eq:BAE}, allows us to find a well-known result for the Bethe roots in the free fermion point 
\begin{align}\label{BetheRoots}
\mathfrak{a}^{\text{sc}}(\lambda, \kappa)=-1 \Longrightarrow \pi\lambda^2 = \frac{1+\kappa}{2}+n, \quad n=0, \pm1, \pm2, \ldots
\end{align}
As is well-known, the ground state eigenvalue of the transfer matrix $T(\lambda)$ corresponds to all Bethe roots being positive. For the spin-zero sector the twist parameter lies in the range $-1<\kappa<1$ , which brings us to the following result:
\begin{align}\label{TA_Explicitly}
& A(\lambda) = \prod_{n=0}^{\infty} \left(1-\frac{\pi\lambda^2}{n+\frac{1+\kappa}{2}}\right)=
	\frac{\Gamma(\frac{1+\kappa}{2})}{\Gamma(-\pi\lambda^2+\frac{1+\kappa}{2})}, \\
& T(\lambda)=\frac{
		2\pi 
				  }
				  {
		\Gamma(\pi\lambda^2+\frac{1+\kappa}{2}) 
		\Gamma(\pi\lambda^2+\frac{1-\kappa}{2})
		          },
\end{align}
which agrees with the result obtained by the authors of \cite{BLZ_1997} for the particular value 0 of their non-universal renormalisation constant $\mathcal{C}$.

Here and everywhere below we omit the upper index sc.

\subsection{A short note on the ODE/IM correspondence}

In recent years, a link between integrable quantum field theories and certain ordinary differential equations in the complex domain was established (for an overview, see \cite{DDT_2007}). Using the ODE/IM correspondence approach we can, for example, rederive results \eqref{BetheRoots}, \eqref{TA_Explicitly} from the previous section. In what follows, however, we will not use all of the tools provided by the method of ODE/IM correspondence. In fact, we only need one specific result from the ODE side: the eigenfunctions of the corresponding Hamiltonian.

The free-fermion point corresponds to the following differential equation
\begin{align}\label{HO_General}
\Ham\psi_n(z)=\left( -\frac{d^2}{dz^2} + z^{2} + \left( \kappa^2-\frac{1}{4} \right) \frac{1}{z^2} \right) \psi_n(z)= E_n\psi_n(z),
\end{align}
which is nothing but the radial part of the three-dimensional harmonic oscillator with the angular momentum $l=\kappa+\frac{1}{2}$. The eigenfunctions of \eqref{HO_General} can be written as
\begin{align}\label{HO_Solution}
\psi_n(z)=z^{\kappa+\frac{1}{2}} e^{-\frac{z^2}{2}} L_n^{(\kappa)}(z^2),
\end{align}
where $L_n^{(\kappa)}(z^2)$ are modified Laguerre polynomials. The spectrum $E_n$ of \eqref{HO_General} is also well known:
\begin{align}\label{HO_Energies}
E_n=4\left(n+\frac{1+\kappa}{2}\right).
\end{align}
Note that it reproduces the Bethe roots \eqref{BetheRoots}, which is one of the results of the ODE/IM correspondence. Notice that the eigenfunctions \eqref{HO_Solution} are not normalised. Let us introduce ``conjugated'' functions
\begin{align}\label{HO_PsiDagger}
\psi_n^\dagger(z) = 2 \frac{n!}{\Gamma(n+\kappa+1)}z^{\kappa+\frac{1}{2}} e^{-\frac{z^2}{2}} L_n^{(\kappa)}(z^2). 
\end{align}
Then, due to the known property of orthogonality of generalised Laguerre polynomials,
\begin{align*}
\int\limits_{0}^{\infty} \psi_m^\dagger(z) \psi_n(z) dz = \delta_{m, n}.
\end{align*}

\subsection{The master function at the free fermion point}
In \cite{BG_2012} the function $\widetilde{\Phi}(\lambda, \mu; \kappa, \kappa', \alpha)$ was defined as a solution of the following equation 
\begin{align}\label{eq:phi-defining-equation}
&\res_{\lambda=\lambda_n}
\left(
\frac{\Delta_\lambda \widetilde{\Phi}(\lambda, \mu; \kappa, \kappa', \alpha) + \psi(\lambda/\mu, \alpha)}{\rho(\lambda; \kappa, \kappa')(1+\mathfrak{a}(\lambda, \kappa))}
+\widetilde{\Phi}(\lambda, \mu; \kappa, \kappa', \alpha)
\right)=0, \\
&\pi\lambda_n^2=n+\frac{1+\kappa}{2}, \quad n\geq0 \nonumber
\end{align}
In the above formula $\Delta_\lambda$ is the finite difference operator
\begin{align}
\Delta_\lambda f(\lambda) = f(q\lambda) - f(q^{-1} \lambda),
\end{align}
the function $\rho(\lambda; \kappa, \kappa')$ is ratio of two eigenvalues of the transfer matrix with different twist parameters:
\begin{align}\label{Def_Rho}
\rho(\lambda; \kappa, \kappa') = \frac{T(\lambda, \kappa')}{T(\lambda, \kappa)} = \frac{\Gamma(\pi\lambda^2+\frac{1+\kappa}{2}) \Gamma(\pi\lambda^2+\frac{1-\kappa}{2})}{\Gamma(\pi\lambda^2+\frac{1+\kappa'}{2}) \Gamma(\pi\lambda^2+\frac{1-\kappa'}{2})}
\end{align}
According to \cite{BG_2012},the function $\Phi(\lambda, \mu; \kappa, \kappa', \alpha)=\widetilde{\Phi}(\lambda, \mu; \kappa, \kappa', \alpha)+\Delta_{\lambda}^{-1}\psi(\lambda/\mu, \alpha)$ allows us to define the function $\omega(\lambda, \mu; \kappa, \kappa', \alpha)$ in a rather simple way:
\begin{align}\label{omega_HHPhi}
	\frac{1}{4}\omega(\lambda, \mu) = H_\lambda H_\mu \Phi(\lambda, \mu).
\end{align}
Here $H_\lambda$ is the difference operator defined as
\begin{align}\label{H_Def}
	H_\lambda f(\lambda) = \frac{1}{1+\mathfrak{a}(\lambda, \kappa)^{-1}} f(q\lambda) + \frac{1}{1+\mathfrak{a}(\lambda, \kappa)} f(q^{-1}\lambda) - \rho(\lambda; \kappa, \kappa') f(\lambda).
\end{align} 

The solution of the equation \eqref{eq:phi-defining-equation} is due to \cite{Smirnov_PC}. We start by noting that the master function has poles at zeros of $Q(\lambda, \kappa)$ and $Q(\mu, \kappa)$. Therefore, to solve the equation (\ref{eq:phi-defining-equation}), we can use the ansatz 
\begin{align}\label{Phi_Ansatz}
\widetilde{\Phi}(\lambda, \mu; \kappa, \kappa', \alpha)=\lambda^\alpha \mu^{2-\alpha} \sum_{m=0}^{\infty} \frac{R_m}{\pi\lambda^2-\left(m+\frac{1+\kappa}{2}\right)},
\end{align}
where the coefficients $R_m$ contain all dependence of the master function on the spectral parameter $\mu$, as well as $\alpha, \kappa$ and $\kappa'$. Plugging the ansatz \eqref{Phi_Ansatz} in \eqref{eq:phi-defining-equation} and using \eqref{TA_Explicitly}, \eqref{psi}, we can easily obtain the equation on the coefficients $R_n$:
\begin{align}\label{Phi_Rn}
R_n + \frac{1}{\rho_n} \frac{\sin\left(\frac{\pi\alpha}{2}\right)}{\pi} \sum_{m=0}^{\infty} \frac{R_m}{m+n+\kappa+1} = \frac{1}{\rho_n} M_n,
\end{align}
where
\begin{align}
& \frac{1}{\rho_n}=\frac{\Gamma\left(n+1+\frac{\kappa+\kappa'}{2}\right) \Gamma\left(n+1+\frac{\kappa-\kappa'}{2}\right)}{\Gamma(n+\kappa+1)\Gamma(n+1)}, \\
& M_n=-\frac{1}{2\irm}\frac{1}{\pi\mu^2 - \left(n+\frac{1+\kappa}{2}\right)}
\end{align}
Let us factorise the multiplier $1/\rho_n$. Namely, set
\begin{align}
\Tilde{R}_n=\frac{\Gamma(n+1)}{\Gamma(n+\frac{\kappa-\kappa'}{2}+1)} R_n.
\end{align}
Then the equation (\ref{Phi_Rn}) becomes
\begin{align}\label{Phi_Rn_Factorised}
&\Tilde{R}_n +
    \frac{\sin(\frac{\pi\alpha}{2})}{\pi} 
        \sum_{m=0}^{\infty} \frac{\Gamma(n+\frac{\kappa+\kappa'}{2}+1)}{\Gamma(n+\kappa+1)} \frac{1}{m+n+\kappa+1} \frac{\Gamma(m+\frac{\kappa-\kappa'}{2}+1)}{\Gamma(m+1)}\Tilde{R}_m \\
&\qquad =\frac{\Gamma(n+\frac{\kappa+\kappa'}{2}+1)}{\Gamma(n+\kappa+1)} M_n. \nonumber
\end{align}
Finally, let us introduce the functions $\Tilde{R}(z)$ and $\Tilde{M}(z)$ via the expansion over the solutions $\psi_n(z)$ of the ODE (\ref{HO_General}):
\begin{align}\label{RM_Def}
\Tilde{R}(z)=\sum_{m=0}^{\infty} \Tilde{R}_m \psi_m(z), 
\quad
\Tilde{M}(z)=\sum_{m=0}^{\infty} \frac{\Gamma(m+\frac{\kappa+\kappa'}{2}+1)}{\Gamma(m+\kappa+1)} M_m \psi_m(z).
\end{align}
With this definition we can rewrite (\ref{Phi_Rn_Factorised}) in an integral form
\begin{align}\label{Phi_RzInt}
\Tilde{R}(z) + \frac{\sin(\frac{\pi\alpha}{2})}{\pi} \int\limits_{0}^{\infty} \Tilde{G}(z, w) \Tilde{R}(w) dw = \Tilde{M}(z),
\end{align}
where 
\begin{align}\label{G_Sum}
\Tilde{G}(z, w) =
 \sum_{m, n=0}^{\infty} \frac{\Gamma(n+\frac{\kappa+\kappa'}{2}+1)}{\Gamma(n+\kappa+1)} \psi_n(z) \frac{1}{m+n+\kappa+1} \frac{\Gamma(m+\frac{\kappa-\kappa'}{2}+1)}{\Gamma(m+1)} \psi_m^\dagger(w)
\end{align}

To proceed, we use the result of the following lemma:
\begin{lemma}\label{lemma:kernel}
The kernel $\tilde{G}(z, w)$ is
\begin{align}\label{eq:G-Def}
\Tilde{G}(z, w) = \left(\frac{z}{w}\right)^{-\kappa'} G(z, w), \quad G(z, w)=2 (zw)^{\frac{1}{2}} \frac{\erm^{-\frac{1}{2}(z^2+w^2)}}{z^2+w^2}.
\end{align}
\end{lemma}
\noindent
We give a proof of this lemma in \Cref{appendix:proof-of-Lemma}.

Let us introduce the function $u(z, \lambda; \kappa, \sigma)$ in the following way:
\begin{align}\label{eq:u-def}
u(z, \lambda; \kappa, \sigma)= 
z^{\kappa+\frac{1}{2}} \erm^{-\frac{z^2}{2}} 
\sum_{m=0}^{\infty} \frac{\Gamma\left(m+\sigma+1\right)}{\Gamma(m+\kappa+1)}
\frac{1}{\pi\lambda^2 - (m+\frac{1+\kappa}{2})} L_m^{(\kappa)}(z^2).
\end{align}
It is easy to notice that $u(z, \lambda; \kappa, \sigma)$ is related to the function $\tilde{M}(z)$:
\begin{align}
u(z, \mu; \kappa, \frac{\kappa+\kappa'}{2})=-2\irm\Tilde{M}(z).
\end{align}
Moreover, the sum involved in the ansatz \eqref{Phi_Ansatz} can be expressed as an integral of $u(z, \lambda; \kappa, \sigma)$
\begin{align}\label{Eq1}
\sum_{m=0}^{\infty} \frac{R_m}{\pi\lambda^2-(m+\frac{1+\kappa}{2})} = 
&\int\limits_0^\infty dz \sum_{m=0}^{\infty} \frac{1}{\pi\lambda^2-(m+\frac{1+\kappa}{2})} \psi_m^\dagger(z) \frac{\Gamma\left(m+\frac{\kappa-\kappa'}{2}+1\right)}{\Gamma(m+1)}\Tilde{R}(z) \\
&=2\int\limits_{0}^{\infty} dz u(z, \lambda; \kappa, \frac{\kappa-\kappa'}{2}) \Tilde{R}(z). \nonumber
\end{align}
At the same time, it follows from \eqref{Phi_RzInt} that
\begin{align}\label{Eq2}
\Tilde{R}(z) = z^{-\kappa'} 
\int\limits_{0}^{\infty}
\left(1+\frac{\sin\frac{\pi\alpha}{2}}{\pi}G\right)^{-1}(z, w) w^{\kappa'} M(w)
\end{align}
Combining \eqref{Eq1} and \eqref{Eq2}, we arrive to the following result
\begin{align}\label{eq:phi-via-inverse-of-G}
\begin{aligned}
&\widetilde{\Phi}(\lambda, \mu; \kappa, \kappa') = \irm\lambda^\alpha \mu^{2-\alpha} \\
&\times \int\limits_0^\infty dz \int\limits_0^\infty dw \cdot
u\left(z, \lambda; \kappa, \frac{\kappa-\kappa'}{2}\right) z^{-\kappa'}
\left(1+\frac{\sin\frac{\pi\alpha}{2}}{\pi}G\right)^{-1}(z, w)
u\left(w, \mu; \kappa, \frac{\kappa+\kappa'}{2}\right) w^{\kappa'}
\end{aligned}
\end{align}

To get the explicit form of the master function we have to find the inverse of the integral operator
in \eqref{eq:phi-via-inverse-of-G}. This calculation is rather technical, so we present it in the \Cref{appendix:inverse}. We summarise the result in the form of the following proposition:
\begin{proposition}\label{proposition:Master-Function}
The master function at the free fermion point is given by a solution of its defining equation \eqref{eq:phi-defining-equation} that can be written in the following form
\begin{align}\label{eq:Phi-Final}
&\widetilde{\Phi}(\lambda, \mu; \kappa, \kappa', \alpha)=
\frac{\irm}{\pi} \lambda^\alpha \mu^{2-\alpha}
\Gamma(-\pi\lambda^2+\frac{1+\kappa}{2}) \Gamma(-\pi\mu^2+\frac{1+\kappa}{2}) \\
&\qquad\times
\int\limits_{0}^{\infty} dk \frac{k \sinh\pi k \cosh\pi k}{\cosh\pi k+\sin\frac{\pi\alpha}{2}} 
\frac{
	F(\lambda, k; \kappa, \kappa') 
	F(\mu, k; \kappa, \kappa')
}
{
	\Gamma(\frac{1+\kappa-\kappa'}{2}+\irm k) 
	\Gamma(\frac{1+\kappa-\kappa'}{2}-\irm k) 
	\Gamma(\frac{1+\kappa+\kappa'}{2}+\irm k) 
	\Gamma(\frac{1+\kappa+\kappa'}{2}-\irm k)
}
\nonumber
\end{align}
where the function $F(\lambda, k; \kappa, \kappa')$ is given by
\begin{align}\label{eq:F-Def}
	F(\lambda, k; \kappa, \kappa')=\int\limits_{-\irm\infty-0}^{+\irm\infty+0} \frac{ds}{2\pi\irm}
	\frac{\Gamma(-s) \Gamma(-s+\frac{\kappa+\kappa'}{2}) \Gamma(-s+\frac{\kappa-\kappa'}{2}) \Gamma(s+ik+\frac{1}{2}) \Gamma(s-ik+\frac{1}{2})}{\Gamma(-s-\pi\lambda^2+\frac{1+\kappa}{2})}.
\end{align}
\end{proposition}
From (\ref{Eq_5}) it is clear that
\begin{align*}
	\widetilde{\Phi}(\lambda, \mu; \kappa, \kappa', \alpha)=\widetilde{\Phi}(\mu, \lambda; \kappa, \kappa', 2-\alpha).
\end{align*}

\section{Correlation functions in the fermionic basis}\label{sec:omega}
As it was discussed above, our main object of interest is the scaling limit of the function $\omega(\lambda, \mu; \kappa, \kappa', \alpha)$. From the asymptotic expansion of the fermionic basis operators \eqref{eq:introduction-fb-asymptotics}, it is clear that
\begin{align}\label{omegasc_Expansion}
	\omega^{\text{sc}}(\lambda, \mu) \simeq \sqrt{\rho^{\text{sc}}(\lambda; \kappa, \kappa')\rho^{\text{sc}}(\mu; \kappa, \kappa')} \sum_{i, j=1}^{\infty} \lambda^{-\frac{2i-1}{\nu}} \mu^{-\frac{2j-1}{\nu}} \omega_{i, j} (\kappa, \kappa', \alpha),
\end{align}
where the coefficients $\omega_{i, j} (\kappa, \kappa', \alpha)$ are to be determined. To do so, we use the relation \eqref{omega_HHPhi} between the master function and $\omega(\lambda, \mu; \kappa, \kappa', \alpha)$. In the scaling limit the operator $H_\lambda$ simplifies. Indeed, from \cite{BLZ_1997} one concludes that for $\lambda^2 \rightarrow \infty$ the function $\mathfrak{a}^{\mathrm{sc}}(\lambda, \kappa)$ rapidly decays in the upper half plane and rapidly grows in the lower half plane. This means that the second term in \eqref{H_Def} does not contribute unless $f(\lambda)$ grows exponentially. Therefore, the relation \eqref{omega_HHPhi} for $\omega^{\mathrm{sc}}(\lambda, \mu; \kappa, \kappa', \alpha)$ becomes
\begin{align}\label{omega_deltadeltaPhi}
\omega^{\text{sc}}(\lambda, \mu) = \delta_\lambda^{-}\delta_\mu^{-}\Phi(\lambda, \mu),
\end{align}
where $\delta^{-}$ is the difference operator defined as
\begin{align}\label{delta_Def}
\delta^{-}_\lambda f(\lambda) = f(q\lambda) - \rho(\lambda; \kappa, \kappa') f(\lambda).
\end{align}

The relation \eqref{omega_deltadeltaPhi} is of utmost importance for us. Using the explicit form \eqref{Eq_5} of the master function, and calculating the asymptotic of the left-hand side of \eqref{omega_deltadeltaPhi} we can compute the coefficients $\omega_{i, j}(\kappa, \kappa', \alpha)$ at the free fermion point.

Before we do that, however, we have to discuss some subtleties involved in taking the scaling limit. It is known (see \cite{BLZ_1997} (3.17) and (3.23)) that for large $\kappa$ the smallest Bethe root behaves as $\lambda_1^2 \sim c(\nu) \kappa^{2\nu}$, where
\begin{align}\label{1_BetheRoot_Smallest}
	c(\nu)=\Gamma(\nu)^{-2}e^\delta\left(\frac{\nu}{2R}\right)^{2\nu},
	\qquad
	\delta=-\nu\log\nu-(1-\nu)\log(1-\nu).
\end{align}
The main technical idea in \cite{BLZ_1997} is to consider the limit
\begin{align}
\lambda^2, \kappa \rightarrow \infty, 
\qquad
\text{ for }
t=c(\nu)^{-1}\frac{\lambda^2}{\kappa^{2\nu}}
\text{ fixed.}
\end{align}
That was done, for example, in \cite{HGS_4, Boos_TBAE, AB_2025} for the case $\kappa=\kappa'$. In order to consider the scaling limit for two different twist parameters $\kappa, \kappa' \rightarrow \infty$, we introduce 
\begin{align}\label{kappa-pm}
\kappa_{+} = \frac{\kappa+\kappa'}{2},
\quad
\kappa_{-}= \frac{\kappa-\kappa'}{2}.
\end{align}
Then the scaling implies that we keep the variables
\begin{align}\label{eq:3-tu-def}
t = \frac{2\pi\lambda^2}{\kappa_{+}}, 
\quad
u = \frac{2\pi\mu^2}{\kappa_{+}}
\end{align}
fixed.

We formulate the main result of this paper in the form of the following proposition:
\begin{proposition}\label{proposition:main-result}
The coefficients $\omega_{i, j}(\kappa, \kappa', \alpha)$ of the asymptotic expansion \eqref{omegasc_Expansion} at the free fermion point are given by
\begin{align}
	\omega_{i, j}(\kappa, \kappa', \alpha)=\widetilde{\omega}_{i, j}(\kappa_{+}, \kappa_{-}, \alpha)+\omega^{(0)}_{i, j}(\kappa_{+}, \kappa_{-}, \alpha),
\end{align}
where $\kappa_{+}$ and $\kappa_{-}$ are defined in \eqref{kappa-pm}. Two terms $\widetilde{\omega}_{i, j}(\kappa_{+}, \kappa_{-}, \alpha)$ and $\omega^{(0)}_{i, j}(\kappa_{+}, \kappa_{-}, \alpha)$ are the following sums:
\begin{align}\label{omega-tilde}
	\begin{aligned}
		\widetilde{\omega}_{i, j}(\kappa_{+}, \kappa_{-}, \alpha) &=  
		-\frac{\irm}{\pi^{i+j}} \frac{\cot\left(\frac{\pi\alpha}{2}\right)}{2(\alpha-1)}\\
		&\times\sum_{k_1, k_2, m=0}^{\infty}(4m+1+\alpha) 
		        \widetilde{K}_{k_1}\left(-i+k_1-\frac{\alpha}{2}; \kappa_{+}, \kappa_{-}\right)
		        S_{m, i-k_1}(\kappa_{+}, \kappa_{-}, \alpha)\\
		&\hspace{1.7cm}\times \widetilde{K}_{k_2}\left(-1-j+k_2+\frac{\alpha}{2}; \kappa_{+}, \kappa_{-}\right)
		         S_{-m-1, j-k_2}(\kappa_{+}, \kappa_{-}, 2-\alpha)
	\end{aligned}
\end{align}
and 
\begin{align}\label{omega-zero}
	\begin{aligned}
		\omega^{(0)}_{i, j}(\kappa_{+}, \kappa_{-}, \alpha) = 2\irm \frac{\cot\left(\frac{\pi\alpha}{2}\right)}{\pi^{i+j}} 
		\Bigg(
		\sum_{m=0}^{\infty} 
		&(-1)^m K^{(0)}_{i-m}(\kappa_{+}, \kappa_{-}) K^{(0)}_{j+m}(\kappa_{+}, \kappa_{-}) \\
		&-\frac{1}{2} K^{(0)}_{i}(\kappa_{+}, \kappa_{-}) K^{(0)}_{j}(\kappa_{+}, \kappa_{-})
		\Bigg)                 
	\end{aligned}
\end{align}
Above the coefficients $S_{m, n}(\kappa_{+}, \kappa_{-}, \alpha)$ are given by
\begin{align}\label{S-Definition}
\begin{aligned}
S_{m, n}(\kappa_{+}, \kappa_{-}, \alpha)&=\frac{\alpha}{(n-2m-1)!}(1-\alpha)_{-n-2m-1}\\
			  &\times \left(1+\frac{\alpha}{2}\right)_{n-1} 
			          \left(1+2m+\frac{\alpha}{2}+\kappa_{-}\right)_{n-2m-1} 
			          \left(1+2m+\frac{\alpha}{2}+\kappa_{+}\right)_{n-2m-1},
\end{aligned}
\end{align}
where $(x)_n$ denotes the Pochhammer symbol (falling factorial), and $\widetilde{K}_i(s; \kappa_{+}, \kappa_{-})$, $K^{(0)}_i(\kappa_{+}, \kappa_{-})$ are the coefficients in the asymptotic expansion of the following combinations of the  gamma functions
\begin{align}
\begin{aligned}
\label{Def_K-Tilde}
&\frac{
		\Gamma\left(\frac{1}{2}-\frac{\kappa_{+}(t-1)}{2} + \frac{\kappa_{-}}{2}\right)
	}
	{
		\Gamma\left(-s+\frac{1}{2}-\frac{\kappa_{+}(t-1)}{2} + \frac{\kappa_{-}}{2}\right)
	} 
\sqrt{
	\frac{
		\Gamma\left(\frac{1}{2}+\frac{\kappa_{+}(t+1)}{2}+\frac{\kappa_{-}}{2}\right)
		\Gamma\left(\frac{1}{2}+\frac{\kappa_{+}(t-1)}{2}-\frac{\kappa_{-}}{2}\right)
		}
		{
		\Gamma\left(\frac{1}{2}+\frac{\kappa_{+}(t+1)}{2}-\frac{\kappa_{-}}{2}\right)
		\Gamma\left(\frac{1}{2}+\frac{\kappa_{+}(t-1)}{2}+\frac{\kappa_{-}}{2}\right)
		}
	}
\\
&\simeq
\left(-\frac{\kappa_{+}t}{2}\right)^{s}
\sum_{i=0}^{\infty}
	\left(\frac{2}{\kappa_{+}t}\right)^i\widetilde{K}_{i}(s; \kappa_{+}, \kappa_{-}) , \\
\end{aligned}
\end{align}
and
\begin{align}
\begin{aligned}
\label{Def_K-Zero}
\sqrt{
	\frac{
		\Gamma\left(\frac{1}{2}+\frac{\kappa_{+}(t+1)}{2}+\frac{\kappa_{-}}{2}\right)
		\Gamma\left(\frac{1}{2}+\frac{\kappa_{+}(t-1)}{2}-\frac{\kappa_{-}}{2}\right)
	}
	{
		\Gamma\left(\frac{1}{2}+\frac{\kappa_{+}(t+1)}{2}-\frac{\kappa_{-}}{2}\right)
		\Gamma\left(\frac{1}{2}+\frac{\kappa_{+}(t-1)}{2}+\frac{\kappa_{-}}{2}\right)
	}
}
\simeq
\sum_{i=0}^{\infty}
\left(\frac{2}{\kappa_{+}t}\right)^i K^{(0)}_{i}(\kappa_{+}, \kappa_{-})
\end{aligned}
\end{align}
Both $\widetilde{K}_i(s; \kappa_{+}, \kappa_{-})$ and $K^{(0)}_i(\kappa_{+}, \kappa_{-})$ are zero for negative indices $i$.
\end{proposition}

We show the derivation of \Cref{proposition:main-result} in \Cref{appendix:omega}. Here we only make some remarks. First, the sums \eqref{omega-tilde} and \eqref{omega-zero} are finite, since they truncate for sufficiently large $m$. Second, the function $S_{m, n}(\kappa_{+}, \kappa_{-}, \alpha)$ can be written in many alternative equivalent forms, for example, via gamma functions. The form \eqref{S-Definition}, however, is the most convenient for computing $\omega_{i, j}$ on a computer.

\subsection{The relation between the Virasoro and fermionic generators}
One of the points of interest for us is to establish the explicit relation between the Fermionic basis and the Virasoro generators. Modulo integrals of motion it was done in \cite{HGS_4, Boos_TBAE} up to level 8. In this section we show how to incorporate the integrals of motion in the free fermion case.

To this end, we take the Virasoro part from (12.4) of \cite{HGS_4} and set $\nu=\frac{1}{2}$. Then we obtain the following:
\begin{align}\label{eq:3-o11-Virasoro}
\omega_{1, 1} 
	= \langle \beta_1^* \gamma_1^* \phi_\alpha(0) \rangle 
	= D_{1}\left(\alpha, \frac{1}{2}\right)D_{1}\left(2-\alpha, \frac{1}{2}\right)
	\left\langle\left(\bbl_{-2} + y^{1, 1}_{1, 1}(\alpha) \bbi_1^2\right) \phi_\alpha(0)\right\rangle
\end{align}
Using the Ward-Takahashi identities (see (6.5) and (6.6) of \cite{HGS_4}) we can compute the expectation value of $\bbl_{-2}\phi_\alpha(0)$ on the right-hand side of \eqref{eq:3-o11-Virasoro}. This gives us
\begin{align}
y^{1, 1}_{1, 1}(\alpha) = \frac{2}{\alpha(\alpha-2)}
\end{align}

Similarly, on the level 4 we have
\begin{align}\label{eq:3-o13-Virasoro}
\begin{aligned}
&\omega_{1, 3} = \langle \beta_1^* \gamma_3^* \phi_\alpha(0) \rangle 
			  = \frac{1}{2} D_{1}\left(\alpha, \frac{1}{2}\right)D_{3}\left(2-\alpha, \frac{1}{2}\right) \\
			  	&\quad
			  	\times\left\langle\left(
			  		\bbl_{-2}^2 + \left(\frac{2c-32}{9}+\frac{2}{3}d_\alpha\right)\bbl_{-4}
			  		+y_{1, 3}^{1, 3}(\alpha) \bbi_1\bbi_3 
			  		+y_{1, 3}^{1, 1, 2}(\alpha)\bbi_1^2 \bbl_{-2}
			  		+y_{1, 3}^{1, 1, 1, 1}(\alpha)\bbi_1^4 
			  	\right)\phi_\alpha(0)\right\rangle,
\end{aligned}
\end{align}
where
\begin{align}
d_\alpha = \frac{\nu(\nu-2)}{\nu-1}(\alpha-1) = \frac{3}{2}(1-\alpha).
\end{align}
Using the Ward-Takahashi identities to calculate the expectation values of even Virasoro generators descendants on the left-hand side of \eqref{eq:3-o13-Virasoro} once again, we come to the following result for the coefficients $y_{1, 3}(\alpha)$:
\begin{subequations}\label{eq:3-y13}
\begin{alignat}{2}
& y_{1, 3}^{1, 3}(\alpha) &&= -\frac{16 \left(\alpha ^3+7 \alpha ^2-17 \alpha -6\right)}{3 (\alpha -6) (\alpha -4) (\alpha -2) \alpha  (\alpha +2)}, \\
& y_{1, 3}^{1, 1, 2}(\alpha) &&= \frac{4 (\alpha +18)}{(\alpha -6) (\alpha -4) (\alpha +2)}, \\
& y_{1, 3}^{1, 1, 1, 1}(\alpha) &&= \frac{4 \left(\alpha ^2-4 \alpha +6\right)}{3 (\alpha -6) (\alpha -4) (\alpha -2) \alpha }
\end{alignat}
\end{subequations}
The coefficients $y_{3, 1}$ in front of the integrals of motion in the expansion of the Fermionic basis generators $\beta_3 \gamma_1$ via the Virasoro generators can be obtained from the above coefficients $y_{1, 3}$ replacing $\alpha$ by $2-\alpha$: 
\begin{subequations}\label{eq:3-y31}
	\begin{alignat}{2}
		& y_{3, 1}^{1, 3}(\alpha) &&= -\frac{16 \left(\alpha ^3-13 \alpha ^2+23 \alpha +4\right)}{3 (\alpha -4) (\alpha -2) \alpha  (\alpha +2) (\alpha +4)}, \\
		& y_{3, 1}^{1, 1, 2}(\alpha) &&= \frac{4 (\alpha -20)}{(\alpha -4) (\alpha +2) (\alpha +4)}, \\
		& y_{3, 1}^{1, 1, 1, 1}(\alpha) &&= \frac{4 \left(\alpha ^2+2\right)}{3 (\alpha -2) \alpha  (\alpha +2) (\alpha +4)}.
	\end{alignat}
\end{subequations}

In order to proceed with higher levels, however, it is not enough to consider the three-point function with the primary fields at the boundary only. We also have to consider the descendants. It was done in \cite{Boos_TBAE, AB_2025} for the case $\kappa=\kappa'$. We hope to incorporate the particle-hole excitations in a future work.

\section{Comparison with known results}
\subsection{The case of equal boundary fields}
The case of equal boundary fields corresponding to the choice $\kappa=\kappa'$ was considered before in \cite{HGS_4} for $\frac{1}{2} \leq \nu \leq 1$. It was done with the help of the Wiener--Hopf factorisation technique. Since it is not applicable to the case $\kappa\neq\kappa'$, we do not discuss it in this work, and instead just present the final result of \cite{HGS_4}. The asymptotic expansion of the function $\omega$ at $\lambda, \mu \rightarrow \infty$ is given by
\begin{align}\label{omega_hgs}
	\omega^{\text{sc}}(\lambda, \mu; \kappa, \kappa, \alpha) \simeq
	\sum_{r, s=1}^{\infty} \frac{1}{r+s-1} D_{2r-1}(\alpha) D_{2s-1}(2-\alpha)
	\lambda^{-\frac{2r-1}{\nu}} \mu^{-\frac{2s-1}{\nu}}
	\Omega_{2r-1, 2s-1}(p, \alpha),
\end{align}
where
\begin{align}
	D_{2n-1}(\alpha)=\frac{1}{\sqrt{i\nu}} \Gamma(\nu)^{-\frac{2n-1}{\nu}} (1-\nu)^{\frac{2n-1}{2}} \frac{1}{(n-1)!}
	\frac{\Gamma\left(\frac{\alpha}{2}+\frac{1}{2\nu}(2n-1)\right)}{\Gamma\left(\frac{\alpha}{2}+\frac{1-\nu}{2\nu}(2n-1)\right)},
\end{align}
and
\begin{align}
	\Omega_{2r-1, 2s-1}(p, \alpha)=
	-\Theta\left(\frac{i(2r-1)}{2\nu}, \frac{i(2s-1)}{2\nu}; p, \alpha \right)
	\left(\frac{r+s-1}{\nu}\right)
	\left(\frac{\sqrt{2}p\nu}{R}\right)^{2r+2s-2}.
\end{align}
The function $\Theta(l, m; \kappa, \alpha)$ has the following asymptotic expansion at $\kappa\rightarrow\infty$
\begin{align}\label{2_Theta_Asymptotics}
	\Theta(l, m; \kappa, \alpha) = \sum_{n=0}^{\infty} \Theta_n(l, m; \alpha) \kappa^{-n}, 
	\qquad
	\Theta_0(l, m; \alpha)=-\frac{i}{l+m}.
\end{align}
It satisfies a recursion equation that allows one to calculate each mode in the above expansion iteratively (see, for example, equation (4.7) of \cite{Boos_TBAE}). The modes $\Theta_n(l, m; \alpha)$ can be found in Appendix C of \cite{Boos_TBAE}. 

Setting $\nu=\frac{1}{2}$ in \eqref{omega_hgs} and $\kappa_{+}=\kappa$, $\kappa_{-}=0$ in \Cref{appendix:modes} we come to the same expressions for the coefficients $\omega_{i, j}$ of the asymptotic expansion of the function $\omega^{(\mathrm{sc})}$. 

\subsection{The reflection relations}
The case of $\frac{1}{2}<\nu<1$ was considered before in \cite{BS_2016}. There the authors managed to incorporate the action of local integrals of motion into the fermionic basis for the sine-Gordon model up to the level 4 with the help of the reflection relations. Let us give a short overview of this work.

Consider the Heisenberg algebra generators $a_k$, $k \in \mathbb{Z} \backslash \lbrace0\rbrace$ satisfying the commutation relations
\begin{align}
[a_k, a_l] = 2k \delta_{k, -l},
\end{align}
as well as the zero-mode $\pi_0$ which is canonical:
\begin{align}
\pi_0 = \frac{\partial}{\irm \partial\phi_0}.
\end{align}
The primary field is identified with the highest vector $|a; 0\rangle$ of the Heisenberg algebra:
\begin{align}
\pi_0 |a; 0\rangle = -\irm a |a; 0\rangle,
\qquad
a_k |a; 0\rangle = 0, \forall k>0.
\end{align}
It is easy to check that the combinations
\begin{subequations}
\begin{align}
\label{eq:4-ln}
&\bbl_n = \frac{1}{4}\sum_{j\neq 0, n} a_j a_{n-j} + \left(\irm(n+1)\frac{Q}{2}+\pi_0\right)a_n,
	\qquad \forall n\neq0, \\
\label{eq:4-l0}
&\bbl_0 = \frac{1}{4}\sum_{j=1}^{\infty} a_{-j} a_{j} + \pi_0(\pi_0+\irm Q),
\end{align} 
\end{subequations}
satisfy the Virasoro algebra with central charge $c=1+6Q^2$:
\begin{align}
[\bbl_m, \bbl_n] = (m-n) \bbl_{n+m}+c\frac{m^3-m}{12}\delta_{m, -n}
\end{align}
and
\begin{align}
\bbl_0 |a; 0\rangle = \Delta |a; 0\rangle,
\qquad \Delta = a(Q-a).
\end{align}

The main point of \cite{BS_2016} is to consider two reflections
\begin{align}
\sigma_1: \alpha \rightarrow -\alpha,
\qquad
\sigma_2: \alpha \rightarrow 2-\alpha.
\end{align}
The Heisenberg descendants of the primary field are invariant under $\sigma_1$ and the Virasoro descendants are invariant under $\sigma_2$, while the integrals of motion are invariant under both. The goal is to find a basis, also invariant under both reflections. 

This problem was solved in \cite{NS_2013}. Let us show how it works at the level 2. Using \eqref{eq:4-ln}, \eqref{eq:4-l0} it is easy to derive that 
\begin{align}
\begin{aligned}
&D_{1}(\alpha, \nu)D_{1}(2-\alpha, \nu)\left(\bbl_{-2}-\frac{\alpha+1}{\alpha}\bbi_1^2\right) |a; 0\rangle \\
&\qquad=D_{1}(\alpha, \nu)D_{1}(2-\alpha, \nu)\frac{1}{4}Q^2 \left(\alpha+\frac{1}{\nu}\right)\left(\alpha-\frac{1-\nu}{\nu}\right)|a; 0\rangle
\end{aligned}
\end{align}
The right hand side of the above identity is invariant under $\sigma_1$. The first term of the left-hand side is invariant under $\sigma_2$, but the second one is not. The idea is to correct this adding $x(\alpha)\bbi_1^2 |a; 0\rangle$ in such a way that it does not spoil the $\sigma_1$-invariance of the right-hand side. This means that the coefficient $x(\alpha)$ should satisfy the following set of equations
\footnote{We take the definition of $D_{2n-1}(\alpha, \nu)$ from \cite{HGS_4}. This definition differs from that used in \cite{BS_2016} by a factor of  $\irm$; as a result, the l.h.s. of \eqref{eq:4-reflection2} appears with the opposite sign compared to \cite{BS_2016}.}
:
\begin{subequations}
\begin{align}
\label{eq:4-reflection1}
&x(\alpha) = x(-\alpha),\\
\label{eq:4-reflection2}
&x(2-\alpha) - x(\alpha) = D_1(\alpha, \nu) D_1(2-\alpha, \nu) \frac{2(1-\alpha)}{\alpha(2-\alpha)}
\end{align}
\end{subequations}
The solution of this equation for $\frac{1}{2}<\nu<1$ is given by (3.6) of \cite{BS_2016} in an integral form. This integral can be calculated in the free fermion point as long as it is understood as a principal value integral. Using the notations of \cite{BS_2016}, we obtain
\begin{subequations}
\begin{alignat}{2}
&A_{1, 1}\left(\alpha, \frac{1}{2}\right) &&= \frac{\irm\alpha\cot\left(\frac{\pi\alpha}{2}\right)}{2\pi^2}, \\
&B_{1, 1}\left(\frac{1}{2}\right) &&= 0,
\end{alignat}
\end{subequations}
so that the overall coefficient in front of $\bbi_1^2$ is 
\begin{align}
-\frac{\alpha+1}{\alpha} - 
 \frac{1}{D_1(\alpha, \nu) D_1(2-\alpha, \nu)}\frac{\irm\alpha\cot\left(\frac{\pi\alpha}{2}\right)}{2\pi^2}
= \frac{2}{\alpha(\alpha-2)},
\end{align}
which is exactly $y_{1, 1}^{1, 1}(\alpha)$

At this point it is important to notice that the solution of \eqref{eq:4-reflection1}, \eqref{eq:4-reflection2} are defined up to arbitrary even and periodic with period 2 function of $\alpha$. In \cite{BS_2016} the minimality assumptions were made which state that, first of all, there are no singularities in the strip $0<\Re(\alpha)<2$ and, second of all, there is no growth for $\Im(\alpha)\rightarrow\pm\infty$. These assumptions were made based on numerical study of the function $\omega(\lambda, \mu)$. The fact that the result at the free fermion point is in agreement with the more general result supports these assumptions.

Now let us discuss level 4. The construction of \cite{BS_2016} becomes cumbersome rather quickly. However, it simplifies greatly at the free fermion point, and the integrals involved in the calculation of the coefficients in front of the integrals of motion can be calculated explicitly. We mention here some typos made in the Appendix of \cite{BS_2016}: some of the coefficients $X_{1, 3}$ and $X_{3, 1}$ are given with wrong overall sign. Namely, one has to perform the following change:
\begin{subequations}
\begin{align}
&X_{1, 3}^{1, 1| 1, 1} \rightarrow -X_{1, 3}^{1, 1| 1, 1}, \\
&X_{1, 3}^{1, 1, 1, 1} \rightarrow -X_{1, 3}^{1, 1, 1, 1}, \\
&X_{3, 1}^{1, 1, 1, 1} \rightarrow -X_{3, 1}^{1, 1, 1, 1}.
\end{align}
\end{subequations}
Taking this into account and setting $\nu=\frac{1}{2}$ in (4.5 - 4.8) of \cite{BS_2016}, one can once again calculate all the integrals taking the principal value for poles on the integration contour. After some algebra, one can obtain \eqref{eq:3-y13}, \eqref{eq:3-y31}.

\section{Conclusions and discussion}
We have presented the explicit form of the master function \eqref{eq:Phi-Final} in the free fermion point. Using its relation to the function $\omega^{\text{sc}}(\lambda, \mu)$, given in the scaling limit by \eqref{omega_deltadeltaPhi}, we computed the asymptotic expansion of $\omega^{\text{sc}}(\lambda, \mu)$ and  derived the explicit expressions for the correlation functions of the fermionic descendants $\langle \beta^*_{2i-1} \gamma^*_{2j-1} \phi_\alpha(0)\rangle$ of the primary field $\phi_\alpha(0)$ (see \Cref{proposition:main-result}).

Our work leaves many directions open for future research. First, we would like to better understand the connection between ODE/IM correspondence and the fermionic basis approach. Deriving the expression \eqref{Eq_5} for the master function, we used the solution \eqref{HO_Solution} of the free fermion ODE \eqref{HO_General}. However, our derivation relied only on the orthogonality of the functions $\psi_n(z)$ and $\psi^\dagger_n(z)$. Generally speaking, any family of orthogonal functions could be used. The particular choice of functions affects the form of the kernel $\tilde{G}(z, w)$. It is remarkable that selecting the solutions of the ODE \eqref{HO_General} leads to such a simple expression for the kernel $\tilde{G}(z, w)$.

Second, we would like to deviate our results outside of the free fermion point. Some of the steps towards this goal were already made in \cite{AB_2025}. In this work we related the results obtained in the free fermion point with the results obtained for $\kappa=\kappa'$ with the help of Wiener--Hopf factorisation technique. We would be interested to see how these results can be used in the case of different boundary conditions as well.

Let us make a small remark related to the last question. In the case of equal boundary conditions, the correspondence between fermionic basis operators and Virasoro generators was established in \cite{HGS_4, Boos_TBAE} modulo integrals of motion up to level 8. Let us denote the combinations of even Virasoro generators as $V_{i, j}(\alpha)$. For example, from \eqref{eq:3-o11-Virasoro} and \eqref{eq:3-o13-Virasoro}, we get
\begin{subequations}
\begin{align}
& V_{1, 1}(\alpha) = \bbl_{-2}, \\
& V_{1, 3}(\alpha) = \bbl_{-2}^2 + \left(\frac{2c-32}{9}+\frac{2}{3}d_\alpha\right)\bbl_{-4}
\end{align}
\end{subequations}
If we want to calculate the expectation value of the same even Virasoro generators, but this time with different left and right states, it is enough to replace $\kappa^2$ by $\kappa_{+}^2+\kappa_{-}^2$ in the final result:
\begin{align}
	\langle \kappa | V_{i, j}(\alpha) | \kappa' \rangle = 
	\langle \kappa | V_{i, j}(\alpha) | \kappa \rangle|_{\kappa^2=\kappa_{+}^2+\kappa_{-}^2} 
\end{align}
We do not know how to prove this statement, but it was checked for all Fermionic basis operators up to level 8.

\section*{Acknowledgements}
SA and HB acknowledge financial support from the DFG in the framework of the research unit FOR 2316 and through project BO 3401/8-1. 

The authors would like to thank F. Smirnov for collaboration at the early stages of this work, which originated from intensive discussions with him. HB thanks F. G\"{o}hmann for his interest and valuable discussions. SA thanks M. Minin for numerous stimulating discussions.

\appendix
\section{Proof of \Cref{lemma:kernel}}\label{appendix:proof-of-Lemma}
In this appendix we present a proof of \Cref{lemma:kernel}. 

Using the definitions \eqref{Phi_Ansatz} and \eqref{HO_PsiDagger} of $\psi_n(z)$ and $\psi_n^\dagger(z)$, we can write $\Tilde{G}(z, w)$ explicitly as
\begin{align}\label{eq:appendix1-GTilde-Explicitly}
	\Tilde{G}(z, w) = 2(zw)^{\kappa+\frac{1}{2}} \erm^{-\frac{1}{2}(z^2+w^2)}
	\sum_{m, n=0}^{\infty} \frac{\Gamma(n+\frac{\kappa+\kappa'}{2}+1)}{\Gamma(n+\kappa+1)}
	\frac{L^{(\kappa)}_m(z^2)	L^{(\kappa)}_m(w^2)	}{m+n+\kappa+1} \frac{\Gamma(m+\frac{\kappa-\kappa'}{2}+1)}{\Gamma(m+1)}
\end{align}
We can rewrite this sum via an integral involving confluent hypergeometric functions $M(a, b; z)$. One of their most important properties is the connection between confluent hypergeometric functions and Laguerre polynomials:
\begin{align}\label{eq:appendix1-HyperGeom-Laguerre-1}
	M\left(a, b, \frac{xy}{x-1}\right)=(1-x)^a \sum_{n=0}^{\infty} \frac{\Gamma(a+n)\Gamma(b)}{\Gamma(a)\Gamma(b+n)} L_n^{(b-1)}(y) x^n.
\end{align}
From this we can easily derive the following identity:
\begin{align}\label{eq:appendix1-HyperGeom-Laguerre-2}
	\frac{\Gamma(\sigma+1)}{\Gamma(\kappa+1)}
		(1-u)^{-\sigma-1} M\left(\sigma+1, \kappa+1, -\frac{xu}{1-u}\right)
		= \sum_{m=0}^{\infty} \frac{\Gamma(\sigma+1+m)}{\Gamma(\kappa+1+m)} u^m L_m^{(\kappa)}(x),
	\quad
	|u|<1.
\end{align}
Using the identity \eqref{eq:appendix1-HyperGeom-Laguerre-2} we rewrite the sum in the expression \eqref{eq:appendix1-GTilde-Explicitly} for the function $\Tilde{G}(z, w)$ as
\begin{align}\label{Lemma1_1}
	&\sum_{m, n=0}^{\infty} L^{(\kappa)}_n(x) \frac{\Gamma(n+\frac{\kappa+\kappa'}{2}+1)}{\Gamma(n+\kappa+1)} \frac{1}{m+n+\kappa+1} L^{(\kappa)}_m(y) \frac{\Gamma(m+\frac{\kappa-\kappa'}{2}+1)}{\Gamma(m+\kappa+1)} \\
	&=\frac{\Gamma(1+\frac{\kappa+\kappa'}{2}) \Gamma(1+\frac{\kappa-\kappa'}{2})}{\Gamma^2(\kappa+1)} \cr
	&\quad\times
		\int\limits_{0}^{1} du \cdot u^\kappa (1-u)^{-\kappa-2} 
		M\left(\frac{\kappa+\kappa'}{2}+1, \kappa+1, -\frac{xu}{1-u}\right) 
		M\left(\frac{\kappa-\kappa'}{2}+1, \kappa+1, -\frac{yu}{1-u}\right).
		\nonumber
\end{align}
	Changing the integration variable to $v=\frac{u}{1-u}$, we continue
\begin{align}\label{Lemma1_2}
\begin{aligned}
	\eqref{Lemma1_1}&=
	\frac{\Gamma(1+\frac{\kappa+\kappa'}{2}) \Gamma(1+\frac{\kappa-\kappa'}{2})}{\Gamma^2(\kappa+1)}\\
	&\quad\times
	\int\limits_{0}^{\infty}dv \cdot v^\kappa 
		M\left(\frac{\kappa+\kappa'}{2}+1, \kappa+1, -xv\right) 
		M\left(\frac{\kappa-\kappa'}{2}+1, \kappa+1, -vy\right).
\end{aligned}
\end{align}
Now we can use one of the Kummer's transformations for confluent hypergeometric functions in order to change the sign of their arguments:
\begin{align*}
	M(a, b, z)=\erm^z M(b-a, b, -z).
\end{align*}
Then
\begin{align}\label{Lemma1_3}
\begin{aligned}
	&\eqref{Lemma1_2}=
	\frac{\Gamma(1+\frac{\kappa-\kappa'}{2})\Gamma(1+\frac{\kappa+\kappa'}{2})}{\Gamma^2(\kappa+1)}
	\int\limits_{0}^{\infty}dv \erm^{-(x+y)v} v^\kappa 
		M\left(\frac{\kappa-\kappa'}{2}, \kappa+1, xv\right) 
		M\left(\frac{\kappa+\kappa'}{2}, \kappa+1, yv\right) \cr
	&=\frac{\Gamma(1+\frac{\kappa-\kappa'}{2})\Gamma(1+\frac{\kappa+\kappa'}{2})}{\Gamma^2(\kappa+1)} x^{-\kappa-1}
	\int\limits_{0}^{\infty} dv \erm^{-\frac{x+y}{x}v} v^\kappa 
	M\left(\frac{\kappa-\kappa'}{2}, \kappa+1, v\right) 
	M\left(\frac{\kappa+\kappa'}{2}, \kappa+1, yv\right) \cr
	&=\frac{\Gamma(1+\frac{\kappa+\kappa'}{2}) \Gamma(1+\frac{\kappa-\kappa'}{2})}{\Gamma(\kappa+1)} 
	x^{-\kappa-1} \left(\frac{x}{y}\right)^{\frac{\kappa-\kappa'}{2}} \left(1+\frac{y}{x}\right)^{-1} 
	{}_2 F_1\left(\frac{\kappa-\kappa'}{2}, \frac{\kappa+\kappa'}{2}; \kappa+1; 1\right) \cr
	&=(xy)^{-\frac{\kappa}{2}} \left(\frac{x}{y}\right)^{-\frac{\kappa'}{2}} \frac{1}{x+y}
\end{aligned}
\end{align}
Multiplying this result for $x=z^2$ and $y=w^2$ by the prefactor coming from \eqref{eq:appendix1-GTilde-Explicitly}, we obtain the statement of the \Cref{lemma:kernel}.

\section{The inverse of the integral operator.}\label{appendix:inverse}
In this appendix we calculate the inverse of the integral operators involved in \eqref{eq:phi-via-inverse-of-G} and derive the \Cref{proposition:Master-Function}.

First of all, the operator $G(z, w)$ given by (\ref{eq:G-Def}) is diagonalisable. Indeed, its eigenfunctions $f(z, k)$ are 
\begin{align}
	&\int\limits_0^\infty dw G(z, w) f(w, k) = \frac{\pi}{\cosh(\pi k)} f(z, k) \\
	&f(z, k)=z^{-\frac{1}{2}} K_{\irm k}\left(\frac{z^2}{2}\right),
\end{align}
where $K_{\irm k}(x)$ is a modified Bessel function of the second kind. The functions $f(z, k)$ are orthogonal for different values of $k$:
\begin{align*}
	\int\limits_0^\infty dz f(z, k) f(z, k') = \frac{\pi^2}{4} \frac{\delta(k-k')}{k \sinh(\pi k)},
\end{align*}
which means that the operator $G(z, w)$ itself admits the following spectral decomposition: 
\begin{align*}
	G(z, w) = \frac{4}{\pi} \int\limits_0^\infty dk \cdot k\tanh(\pi k) f(z, k) f(w, k).
\end{align*}
After that the inverse can be written as
\begin{align*}
	\left(1+\frac{\sin\frac{\pi\alpha}{2}}{\pi}G\right)^{-1}(z, w) = 
	\frac{4}{\pi^2}
	\int\limits_0^\infty dk \frac{k\sinh\pi k \cosh\pi k}{\cosh\pi k + \sin\frac{\pi\alpha}{2}}
	\frac{1}{(zw)^\frac{1}{2}} K_{ik}(\frac{1}{2}z^2) K_{ik}(\frac{1}{2}w^2)
\end{align*}

The function $u(z, \lambda; \kappa, \sigma)$ introduced in \eqref{eq:u-def} admits the following integral representation for $0<\sigma<1$ and $\pi\lambda^2-\frac{1+\kappa}{2}<-\sigma$ 
\begin{align*}
	u(z, \lambda; \kappa, \sigma) = -z^{\kappa+\frac{1}{2}} \erm^{-\frac{z^2}{2}} \frac{\irm}{2\Gamma(-\pi\lambda^2 + \frac{1+\kappa}{2} - \sigma)}
	\int\limits_\gamma ds \frac{\Gamma(-s) \Gamma(s-\pi\lambda^2+\frac{1+\kappa}{2})}{\Gamma(s+\kappa+1) \sin\pi(s+\sigma)} z^{2s},
\end{align*}
where the contour $\gamma$ goes from $-\irm\infty-0-\sigma$ to $+\irm\infty-0-\sigma$. We also can write the Mellin transform for the Bessel function
\begin{align*}
	f(z, k) = \pi^{-3/2} \erm^{-\frac{z^2}{2}} z^{2\irm k -\frac{1}{2}} \frac{\cosh(\pi k)}{2\irm} \int\limits_{-\irm\infty-0}^{+\irm\infty-0} dt \Gamma(\frac{1}{2}+\irm k+t) \Gamma(-t) \Gamma(-2\irm k-t) z^{2t}
\end{align*}
Then we can calculate the following integral present in (\ref{eq:phi-via-inverse-of-G}):
\begin{align}\label{Phi_BigInt}
	&-\int\limits_0^\infty dz z^{-\kappa'} u(z, \lambda; \kappa, \frac{\kappa-\kappa'}{2}) f(z, k)
	=\int\limits_0^\infty dz 
		\cdot z^{\kappa-\kappa'+ 2\irm k} \erm^{-z^2} 
		\frac{\cosh(\pi k)}{8\pi^{3/2} \Gamma(-\pi\lambda^2 + \frac{1+\kappa'}{2})}\\
	&\qquad \times 
		\int\limits_\gamma ds 
			z^{2s} \frac{\Gamma(-s) \Gamma(s-\pi\lambda^2+\frac{1+\kappa}{2})}{\Gamma(s+\kappa+1) \sin\pi(s+\frac{\kappa-\kappa'}{2})}
		\int\limits_{-\irm\infty-0}^{+\irm\infty-0} dt 
			z^{2t} \Gamma(\frac{1}{2}+\irm k+t) \Gamma(-t) \Gamma(-2\irm k-t) \cr
	&=\frac{\cosh(\pi k)}{8 \pi^{3/2} \Gamma(-\pi\lambda^2 + \frac{1+\kappa'}{2})}
	\int\limits_\gamma ds \int\limits_{-\irm\infty-0}^{+\irm\infty-0} dt \frac{\Gamma(-s) \Gamma(s-\pi\lambda^2+\frac{1+\kappa}{2})}{\Gamma(s+\kappa+1) \sin\pi(s+\frac{\kappa-\kappa'}{2})} \cr
	&\qquad\times \Gamma(\frac{1}{2}+\irm k+t) \Gamma(s+\frac{\kappa-\kappa'}{2}+\frac{1}{2}+\irm k+t) \Gamma(-t) \Gamma(-2\irm k-t) \cr
	&=\frac{\irm \cosh(\pi k)}{4 \pi^{1/2} \Gamma(-\pi\lambda^2 + \frac{1+\kappa'}{2})}
	\int\limits_\gamma ds \frac{\Gamma(-s) \Gamma(s-\pi\lambda^2+\frac{1+\kappa}{2})}{\Gamma(s+\kappa+1) \sin\pi(s+\frac{\kappa-\kappa'}{2})} \cr
	&\qquad \times
	\frac{\Gamma(\frac{1}{2}+\irm k) \Gamma(\frac{1}{2}-\irm k) \Gamma(\frac{1}{2}+s+\irm k+\frac{\kappa-\kappa'}{2}) \Gamma(\frac{1}{2}+s-\irm k+\frac{\kappa-\kappa'}{2})}{\Gamma(1+s+\frac{\kappa-\kappa'}{2})} \cr
	&=\frac{\irm\pi^{1/2}}{4 \Gamma(-\pi\lambda^2 + \frac{1+\kappa'}{2})}\int\limits_\gamma ds 
	\frac{\Gamma(-s) \Gamma(s-\pi\lambda^2+\frac{1+\kappa}{2})}{\Gamma(s+\kappa+1) \sin\pi(s+\frac{\kappa-\kappa'}{2})}
	\frac{\Gamma(\frac{1}{2}+s+\irm k+\frac{\kappa-\kappa'}{2}) \Gamma(\frac{1}{2}+s-\irm k+\frac{\kappa-\kappa'}{2})}{\Gamma(1+s+\frac{\kappa-\kappa'}{2})} \cr
	&=\frac{\irm\pi^{1/2}}{4 \Gamma(-\pi\lambda^2 + \frac{1+\kappa'}{2})}
		\int\limits_{-\irm\infty-0}^{+\irm\infty-0} ds 
	\frac{\Gamma(-s + \frac{\kappa-\kappa'}{2}) \Gamma(s-\pi\lambda^2+\frac{1+\kappa'}{2}) \Gamma(\frac{1}{2}+s+\irm k)\Gamma(\frac{1}{2}+s-\irm k)}{\Gamma(s+\frac{\kappa+\kappa'}{2}+1) \Gamma(s+1) \sin(\pi s)} \cr
	&=-\frac{\irm}{4\pi^{1/2} \Gamma(-\pi\lambda^2 + \frac{1+\kappa'}{2})} 
		\int\limits_{-\irm\infty-0}^{+\irm\infty-0}  ds 
	\frac{\Gamma(-s) \Gamma(-s+\frac{\kappa-\kappa'}{2}) \Gamma(s-\pi\lambda^2+\frac{1+\kappa'}{2}) \Gamma(\frac{1}{2}+s+\irm k)\Gamma(\frac{1}{2}+s-\irm k)}{\Gamma(s+\frac{\kappa+\kappa'}{2}+1)} \nonumber
\end{align}

The integral over $w$ in (\ref{eq:phi-via-inverse-of-G}) can be calculated the same way. We, therefore, obtain the following result:
\begin{align}\label{Phi_Final2}
	&\widetilde{\Phi}(\lambda, \mu; \kappa, \kappa', \alpha)=\\
	&-\irm\frac{\lambda^\alpha \mu^{2-\alpha}}{4\pi^3 \Gamma(-\pi\lambda^2+\frac{1+\kappa'}{2}) \Gamma(-\pi\mu^2+\frac{1-\kappa'}{2})}
	\int\limits_{0}^{\infty} dk \frac{k \sinh\pi k \cosh\pi k}{\cosh\pi k+\sin\frac{\pi\alpha}{2}} F_{+}(\lambda, k; \kappa, -\kappa') F_{+}(\mu, k; \kappa, \kappa'), \nonumber
\end{align}
where
\begin{align}\label{Phi_FPlus}
	F_{+}(\lambda, k; \kappa, \kappa')=\int\limits_{-\irm\infty-0}^{+\irm\infty-0}ds 
	\frac{\Gamma(-s) \Gamma(-s+\frac{\kappa+\kappa'}{2}) \Gamma(s-\pi\lambda^2+\frac{1-\kappa'}{2}) \Gamma(s+\frac{1}{2}+\irm k) \Gamma(s+\frac{1}{2}-\irm k)}{\Gamma(s+1+\frac{\kappa-\kappa'}{2})}
\end{align}

The formula (\ref{Phi_Final2}) can be brought to a form that is explicitly $\alpha \leftrightarrow 2-\alpha$ symmetric. To do so we apply the first Barnes lemma to the function $F_{+}(\lambda, k; \kappa, \kappa')$. Then
\begin{align}
	&\frac{\Gamma(s+\frac{1}{2}+\irm k) \Gamma(s+\frac{1}{2}-\irm k)}{\Gamma(s+1+\frac{\kappa-\kappa'}{2})}=\\
	&\frac{1}{\Gamma(\irm k+\frac{1+\kappa-\kappa'}{2}) \Gamma(-\irm k+\frac{1+\kappa-\kappa'}{2})}
	\int\limits_{-\irm\infty}^{+\irm\infty} \frac{dw}{2\pi\irm} \Gamma(w+\irm k+\frac{1}{2}) \Gamma(w-\irm k+\frac{1}{2}) \Gamma(-w+s) \Gamma(-w+\frac{\kappa-\kappa'}{2}). \nonumber
\end{align}
Hence,
\begin{align*}
	&F_{+}(\lambda, k; \kappa, \kappa')=
	\int\limits_{-\irm\infty-0}^{+\irm\infty-0}ds \int\limits_{-\irm\infty-0}^{+\irm\infty-0}\frac{dw}{2\pi \irm}
	\frac{\Gamma(-s) \Gamma(-s+\frac{\kappa+\kappa'}{2}) \Gamma(s-\pi\lambda^2+\frac{1-\kappa'}{2})}{\Gamma(\irm k+\frac{1+\kappa-\kappa'}{2}) \Gamma(-\irm k+\frac{1+\kappa-\kappa'}{2})}\times\cr
	&\hspace{2.3cm} \Gamma(w+\irm k+\frac{1}{2}) \Gamma(w-\irm k+\frac{1}{2}) \Gamma(-w+s) \Gamma(-w+\frac{\kappa-\kappa'}{2}) \nonumber
\end{align*}
The integral with respect to $s$ can be calculated also with the help of the first Barnes lemma:
\begin{align}
\begin{aligned}
	&\int\limits_{-\irm\infty-0}^{+\irm\infty-0}\frac{ds}{2\pi i} \Gamma(-s) \Gamma(-s+\frac{\kappa+\kappa'}{2}) \Gamma(s-\pi\lambda^2+\frac{1-\kappa'}{2}) \Gamma(-w+s)  \\
	&=\frac{\Gamma(-w) \Gamma(-w+\frac{\kappa+\kappa'}{2}) \Gamma(-\pi\lambda^2+\frac{1-\kappa'}{2}) \Gamma(-\pi\lambda^2+\frac{1+\kappa}{2})}{\Gamma(-w-\pi\lambda^2+\frac{1+\kappa}{2})}
\end{aligned}
\end{align}
Therefore, the final answer reads 
\begin{align}
\begin{aligned}
	&F_{+}(\lambda, k; \kappa, \kappa')=\frac{\Gamma(-\pi\lambda^2+\frac{1-\kappa'}{2}) \Gamma(-\pi\lambda^2+\frac{1+\kappa}{2})}{\Gamma(\irm k+\frac{1+\kappa-\kappa'}{2}) \Gamma(-\irm k+\frac{1+\kappa-\kappa'}{2})}\\
	&\times\int\limits_{-\irm\infty-0}^{+\irm\infty-0} dw \frac{\Gamma(-w) \Gamma(-w+\frac{\kappa+\kappa'}{2}) \Gamma(-w+\frac{\kappa-\kappa'}{2}) \Gamma(w+\irm k+\frac{1}{2}) \Gamma(w-\irm k+\frac{1}{2})}{\Gamma(-w-\pi\lambda^2+\frac{1+\kappa}{2})}
\end{aligned}
\end{align}
Let us denote the integral in the last line as new function $F(\lambda, k; \kappa, \kappa')$:
\begin{align}\label{Eq_4}
	F(\lambda, k; \kappa, \kappa')=\int\limits_{-\irm\infty-0}^{+\irm\infty+0} \frac{ds}{2\pi\irm}
	 \frac{\Gamma(-s) \Gamma(-s+\frac{\kappa+\kappa'}{2}) \Gamma(-s+\frac{\kappa-\kappa'}{2}) \Gamma(s+ik+\frac{1}{2}) \Gamma(s-ik+\frac{1}{2})}{\Gamma(-s-\pi\lambda^2+\frac{1+\kappa}{2})}.
\end{align}
Then (\ref{Phi_Final2}) can be rewritten as
\begin{align}\label{Eq_5}
\begin{aligned}
	&\widetilde{\Phi}(\lambda, \mu; \kappa, \kappa', \alpha)=
	\frac{\irm}{\pi} \lambda^\alpha \mu^{2-\alpha}
	\Gamma(-\pi\lambda^2+\frac{1+\kappa}{2}) \Gamma(-\pi\mu^2+\frac{1+\kappa}{2}) \\
	&\quad\times
	\int\limits_{0}^{\infty} dk \frac{k \sinh\pi k \cosh\pi k}{\cosh\pi k+\sin\frac{\pi\alpha}{2}} 
	\frac{
			F(\lambda, k; \kappa, \kappa') 
			F(\mu, k; \kappa, \kappa')
		 }
		 {
		 	\Gamma(\frac{1+\kappa-\kappa'}{2}+\irm k) 
		 	\Gamma(\frac{1+\kappa-\kappa'}{2}-\irm k) 
		 	\Gamma(\frac{1+\kappa+\kappa'}{2}+\irm k) 
		 	\Gamma(\frac{1+\kappa+\kappa'}{2}-\irm k)
		 }
\end{aligned}
\end{align}
This gives us the result of the \Cref{proposition:Master-Function}

\section{The function $\omega^{(\mathrm{sc})}$}\label{appendix:omega}
In this appendix we show the derivation of the \Cref{proposition:main-result}. 

We start with the relation \eqref{omega_deltadeltaPhi}
\begin{align}\label{eq:C-omega-via-Phi}
\omega^{\text{sc}}(\lambda, \mu) 
= \delta_\lambda^{-}\delta_\mu^{-}\Phi(\lambda, \mu) 
= \delta_\lambda^{-}\delta_\mu^{-}\widetilde{\Phi}(\lambda, \mu) + \delta_\lambda^{-}\delta_\mu^{-}\Delta^{-1}\psi\left(\frac{\lambda}{\mu}, \alpha\right).
\end{align}
We calculate the asymptotic expansion of each term separately. Since we are, however, interested in the coefficients $\omega_{i, j}$ of the expansion \eqref{omegasc_Expansion}, our aim is to calculate the asymptotic expansion of 
\begin{align}\label{omega_PhiWithQ}
\begin{aligned}
\sum_{i, j=1}^{\infty} 
	\lambda^{-\frac{2i-1}{\nu}} \mu^{-\frac{2j-1}{\nu}} \omega_{i, j} (\kappa, \kappa', \alpha) &=
		\frac{1}{\sqrt{\rho(\lambda)\rho(\mu)}}\Phi(q\lambda, q\mu)
		-\sqrt{\frac{\rho(\lambda)}{\rho(\mu)}}\Phi(\lambda, q\mu) \\
	  & - \sqrt{\frac{\rho(\mu)}{\rho(\lambda)}}\Phi(q\lambda, \mu) 
	  	+ \sqrt{\rho(\lambda)\rho(\mu)}\Phi(\lambda, \mu). 
\end{aligned}
\end{align}

In variables $t$, $u$ given by \eqref{eq:3-tu-def}, the expression for the function $\widetilde{\Phi}(t, u)$ reads
\begin{align}\label{PhiPrime_tu}
	&\widetilde{\Phi}(t, u; \kappa, \kappa', \alpha)=
	\frac{\irm \kappa_{+}}{2\pi^2} t^{\frac{\alpha}{2}} u^{1-\frac{\alpha}{2}}
	\Gamma\left(\frac{1}{2} - \frac{\kappa_{+}t}{2} + \frac{\kappa_{+}+\kappa_{-}}{2}\right)
	\Gamma\left(\frac{1}{2} - \frac{\kappa_{+}u}{2} + \frac{\kappa_{+}+\kappa_{-}}{2}\right) \cr
	&\int\limits_{0}^{\infty} dk \frac{k \sinh\pi k \cosh\pi k}{\cosh\pi k+\sin\frac{\pi\alpha}{2}} 
	\frac{F(t, k; \kappa_{+}, \kappa_{-}) F(u, k; \kappa_{+}, \kappa_{-})}{\Gamma(\frac{1}{2}+\kappa_{+}+\irm k) \Gamma(\frac{1}{2}+\kappa_{+}-\irm k) \Gamma(\frac{1}{2}+\kappa_{-}+\irm k) \Gamma(\frac{1}{2}+\kappa_{-}-\irm k)}, 
\end{align}
where now the function $F$ is 
\begin{align}\label{F_tu}
	F(t, k; \kappa, \kappa')=\int\limits_{-\irm\infty-0}^{+\irm\infty-0} \frac{ds}{2\pi\irm} \frac{\Gamma(-s) \Gamma(-s+\kappa_{+}) \Gamma(-s+\kappa_{-}) \Gamma(s+\irm k+\frac{1}{2}) \Gamma(s-\irm k+\frac{1}{2})}{\Gamma\left(-s+\frac{1}{2}-\frac{\kappa_{+}t}{2}+\frac{\kappa_{+}+\kappa_{-}}{2}\right)}.
\end{align}
In \eqref{eq:C-omega-via-Phi}, we combine all functions that depend on $t$ or $u$ introducing the function $\widetilde{K}(t; s; \kappa_{+}, \kappa_{-})$ as the following series:
\begin{align}
	\begin{aligned}
		\label{Def_K-Tilde}
		\frac{
			\Gamma\left(\frac{1}{2}-\frac{\kappa_{+}(t-1)}{2} + \frac{\kappa_{-}}{2}\right)
		}
		{
			\Gamma\left(-s+\frac{1}{2}-\frac{\kappa_{+}(t-1)}{2} + \frac{\kappa_{-}}{2}\right)
		} 
		\sqrt{\rho(t; \kappa_{+}, \kappa_{-})}
		&\simeq
		\left(-\frac{\kappa_{+}t}{2}\right)^{s}
		\sum_{i=0}^{\infty}
		\left(\frac{2}{\kappa_{+}t}\right)^i\widetilde{K}_{i}(s; \kappa_{+}, \kappa_{-}) , \\
		&=\left(-\frac{\kappa_{+}t}{2}\right)^{s} \widetilde{K}(t; s; \kappa_{+}, \kappa_{-}).
	\end{aligned}
\end{align}
Then
\begin{align}\label{eq:C-deltadeltaPhiTilde-explicitly}
&\frac{1}{\sqrt{\rho(t)\rho(u)}}\delta^-_t\delta^-_u\widetilde{\Phi}(t, u) = \\
&
		=\int\limits_{0}^{\infty} dk \frac{k \sinh\pi k \cosh\pi k}{\cosh\pi k+\sin\frac{\pi\alpha}{2}} 
		\frac{1}{
			\Gamma(\frac{1}{2}+\kappa_{+}+\irm k)
			\Gamma(\frac{1}{2}+\kappa_{+}-\irm k)
			\Gamma(\frac{1}{2}+\kappa_{-}+\irm k)
			\Gamma(\frac{1}{2}+\kappa_{-}-\irm k)
			}\cr
		&\qquad
		\times\int\limits_{-\irm\infty-0}^{+\irm\infty-0} ds_1
		\Gamma(-s_1)\Gamma(-s_1+\kappa_{+})\Gamma(-s_1+\kappa_{-})\Gamma(s_1+\irm k+\frac{1}{2})\Gamma(s_1-\irm k+\frac{1}{2})\cr
		&\qquad
		\times \int\limits_{-\irm\infty-0}^{+\irm\infty-0} ds_2\Gamma(-s_2)\Gamma(-s_2+\kappa_{+})\Gamma(-s_2+\kappa_{-})\Gamma(s_2+\irm k+\frac{1}{2})\Gamma(s_2-\irm k+\frac{1}{2}) \cr
		&\qquad\qquad
		\times\left(\frac{\kappa_{+}t}{2}\right)^{s_1}
			  \left(\frac{\kappa_{+}u}{2}\right)^{s_2} 
		      t^{\frac{\alpha}{2}} u^{1-\frac{\alpha}{2}} 
		\Big\lbrace
		-\widetilde{K}(-t; s_1; \kappa_{+}, \kappa_{-})\widetilde{K}(-u; s_2; \kappa_{+}, \kappa_{-}) \cr
		&\qquad\qquad\qquad
			-(-1)^{s_2}\erm^{\frac{\irm\pi\alpha}{2}} 
			\widetilde{K}(-t; s_1; \kappa_{+}, \kappa_{-})\widetilde{K}(u; s_2; \kappa_{+}, \kappa_{-}) \cr
		&\qquad\qquad\qquad
		+(-1)^{s_1}\erm^{-\frac{\irm\pi\alpha}{2}} 
			\widetilde{K}(t; s_1; \kappa_{+}, \kappa_{-})\widetilde{K}(-u; s_2; \kappa_{+}, \kappa_{-})\cr
		&\qquad\qquad\qquad
		+(-1)^{s_1+s_2} 
			\widetilde{K}(t; s_1; \kappa_{+}, \kappa_{-})\widetilde{K}(u; s_2; \kappa_{+}, \kappa_{-})
		\Big\rbrace.
		\nonumber
\end{align}
We calculate the integrals in \eqref{eq:C-deltadeltaPhiTilde-explicitly} using the residue theorem. Picking the right residues, we make sure that $t$ and $u$ enter the final expression in the desired power, thus obtaining the coefficient $\omega_{i, j}$. Let us explain how we do this. Since $t, u \gg 1$, the integration contour in the integrals with respect to $s_1$ and $s_2$ can be closed to the left. In the left half-plane, there are two families of residues
\begin{subequations}
\begin{align}
&s_{1, 2} = -\frac{1}{2} - \irm k -n, n=0, 1, 2, \ldots \\
&s_{1, 2} = -\frac{1}{2} + \irm k -n, n=0, 1, 2, \ldots 
\end{align}
\end{subequations}
In order to calculate the integral with respect to $k$, we notice that the integrand is an even function of $k$, and the integration can be, therefore, performed over the entire real axis. If we assume that $t>u$, the integration contour can be closed in the upper half-plane. This choice does not restrict us at all since $\omega(t, u; \alpha) = \omega(u, t; 2-\alpha)$. In the upper half-plane there are the following poles with respect to $k$:
\begin{subequations}
\begin{align}
&k=\frac{\irm \alpha}{2} + \frac{\irm}{2} + 2\irm m, m=0, 1, 2, \ldots, \\
&k=-\frac{\irm \alpha}{2} + \frac{3\irm}{2} + 2\irm m, m=0, 1, 2, \ldots.
\end{align}
\end{subequations}

Writing the integrals in \eqref{eq:C-deltadeltaPhiTilde-explicitly} as the sums over the above mentioned residues, we can derive \eqref{omega-tilde}. Let us, however, give some comments here. First of all, the expansion \eqref{omegasc_Expansion} of the function $\omega(\lambda, \mu)$ goes in odd powers of $\lambda^{2}$, $\mu^{2}$. In our case, only two sums out of 8 contribute to this expansion (indeed, one can make sure that the other 6 sums vanish due to simple trigonometric identities). This simplifies the calculations. The equation \eqref{eq:C-deltadeltaPhiTilde-explicitly} with this in mind becomes
\begin{align}
\begin{aligned}
&\eqref{eq:C-deltadeltaPhiTilde-explicitly} = \\
&
\sum\limits_{\substack{m, n_1, n_2=0, \\ k_1, k_2=0}}^{\infty} 
	\frac{
		(4m+1+\alpha)\sin\frac{\pi\alpha}{2}
		 }{
		 \Gamma(-2m-\frac{\alpha}{2}+\kappa_{+})
		 \Gamma(2m+1+\frac{\alpha}{2}+\kappa_{+})
		 \Gamma(-2m-\frac{\alpha}{2}+\kappa_{-})
		 \Gamma(2m+1+\frac{\alpha}{2}+\kappa_{-})
		 }\\
&\qquad\times
	\frac{-\irm}{2\pi^2 n_1!} \left(1+(-1)^{n_1+k_1}\erm^{\irm\pi\alpha}\right)
				   \Gamma(-1-4m-n_1-\alpha)
				   \Gamma\left(1+2m+n_1+\frac{\alpha}{2}\right)\\
&\qquad\qquad\times
				   \Gamma\left(1+2m+n_1+\frac{\alpha}{2}+\kappa_{+}\right)
				   \Gamma\left(1+2m+n_1+\frac{\alpha}{2}+\kappa_{-}\right)\\
&\qquad\times
	\frac{1}{n_2!}\left(1+(-1)^{n_2+k_2}\erm^{-\irm\pi\alpha}\right)
			       \Gamma(1+4m-n_2+\alpha)
				   \Gamma\left(-2m+n_2-\frac{\alpha}{2}\right)\\
&\qquad\qquad	\times
			       \Gamma\left(-2m+n_2-\frac{\alpha}{2}+\kappa_{+}\right)
				   \Gamma\left(1+2m+n_1+\frac{\alpha}{2}+\kappa_{-}\right)\\
& \qquad\times
  \widetilde{K}_{k_1}(-1-2m-n_1-\frac{\alpha}{2}; \kappa_{+}, \kappa_{-})
	\widetilde{K}_{k_2}(2m-n_2+\frac{\alpha}{2}; \kappa_{+}, \kappa_{-}).
\end{aligned}
\end{align}
In the above sum the only contribution to $\omega_{i, j}$ comes from $k_1=-1-2m-n_1+i \geq 0$, $k_2=1+2m-n_2+j \geq 0$. After some simple algebra, one can derive \eqref{omega-tilde}.

The final piece of the puzzle is the asymptotic expansion of the function
\begin{align}\label{eq:C-omega0-def}
	\omega_0(\lambda, \mu; \kappa, \kappa', \alpha) = 
	\delta^-_\lambda \delta^-_\mu \Delta^{-1}_{\lambda} \psi\left(\lambda/\mu, \alpha\right).
\end{align}
The inverse of the operator $\Delta^{-1}$ can be defined in multiple ways. For us, the most convenient form is the one given in \cite{JMS_Book1} by
\begin{align}
\Delta^{-1}_\lambda \psi(\lambda, \alpha) = 
	\text{p.v.} \int\limits_{-\infty}^{+\infty} dk \zeta^{2\irm k} 
		\frac{
			\irm\coth\pi\left(k+\frac{\irm\alpha}{2}\right)
			}
			{
			4\sinh\pi k
			},
\end{align}
where the principal value is taken at the pole $k=0$. Performing the same calculations as above, it is easy to derive that the asymptotic expansion of \eqref{eq:C-omega0-def} is given by \eqref{omega-zero}.

\section{Modes $\omega_{i, j}$}\label{appendix:modes}
In this section we give the explicit results for first several coefficients $\omega_{i, j}$ 

\begin{align*}
\omega_{1, 1}&=\frac{\irm \cot\left(\frac{\pi\alpha}{2}\right)}{192 \pi ^2 (\alpha-1)}
\left(
(-2+\alpha ) \alpha  \left(4-2 \alpha +\alpha ^2\right)-12 (-2+\alpha ) \alpha  \kappa _-^2-12 (-2+\alpha ) \alpha  \kappa _+^2-48 \kappa _-^2 \kappa _+^2
\right)\\
\omega_{1, 3}&=
	\frac{\irm\cot \left(\frac{\pi  \alpha }{2}\right)}{737280(\alpha-1)(\alpha-3) \pi ^4}
    \Big(
	(-6+\alpha ) (-4+\alpha ) (-2+\alpha ) \alpha  \left(336-160 \alpha +148 \alpha ^2-44 \alpha ^3+5 \alpha ^4\right)\\
	&-120 (-6+\alpha ) (-4+\alpha ) (-2+\alpha ) \alpha  \left(12-2 \alpha +\alpha ^2\right)
	\kappa _-^2+720 (-6+\alpha ) (-4+\alpha ) (-2+\alpha ) \alpha  \kappa _-^4\\
	&-120 (-6+\alpha ) (-4+\alpha ) (-2+\alpha ) \alpha  \left(12-2 \alpha +\alpha ^2\right) \kappa _+^2
	 +720 (-6+\alpha ) (-4+\alpha ) (-2+\alpha ) \alpha  \kappa _+^4\\
	&+960 \left(-48-148 \alpha
	+126 \alpha ^2-24 \alpha ^3+\alpha ^4\right) \kappa _-^2 \kappa _+^2-1920 (2+\alpha ) (-12+5 \alpha ) \kappa _-^4 \kappa _+^2\\
	&-1920
	(2+\alpha ) (-12+5 \alpha ) \kappa _-^2 \kappa _+^4+3840 \left(6-4 \alpha +\alpha ^2\right) \kappa _-^4 \kappa _+^4
	\Big)\\
\omega_{3, 1}&=
	-\frac{\irm\cot\left(\frac{\pi\alpha}{2}\right)}{737280(\alpha^2-1)\pi^4}
	\Big(
	(-2+\alpha ) \alpha  (2+\alpha ) (4+\alpha ) \left(336-64 \alpha +4 \alpha ^2+4 \alpha ^3+5 \alpha ^4\right)\\
	&-120 (-2+\alpha ) \alpha  (2+\alpha ) (4+\alpha ) \left(12-2 \alpha +\alpha ^2\right) \kappa
	_-^2+720 (-2+\alpha ) \alpha  (2+\alpha ) (4+\alpha ) \kappa _-^4\\
	&-120 (-2+\alpha ) \alpha  (2+\alpha ) (4+\alpha ) \left(12-2 \alpha +\alpha ^2\right) \kappa _+^2+
	+720 (-2+\alpha ) \alpha  (2+\alpha ) (4+\alpha ) \kappa _+^4\\
	&+960 \left(-16-100 \alpha +6 \alpha
	^2+16 \alpha ^3+\alpha ^4\right) \kappa _-^2 \kappa _+^2-1920 (-4+\alpha ) (2+5 \alpha ) \kappa _-^4 \kappa _+^2\\
	&-1920 (-4+\alpha ) (2+5
	\alpha ) \kappa _-^2 \kappa _+^4+3840 \left(2+\alpha ^2\right) \kappa _-^4 \kappa _+^4
	\Big)\\
\omega_{1, 5} &= 
	\frac{\irm\cot \left(\frac{\pi  \alpha }{2}\right)}{5945425920(\alpha-1)(\alpha-3)(\alpha-5) \pi ^6}
	\Big(
	(-10+\alpha ) (-8+\alpha ) (-6+\alpha ) (-4+\alpha ) (-2+\alpha ) \alpha  \\
	&\qquad\times \left(89280-38496 \alpha +43808 \alpha ^2-21000 \alpha ^3+5648 \alpha ^4-714 \alpha ^5+35 \alpha ^6\right)\\
	&-252 (-10+\alpha )
	(-8+\alpha ) (-6+\alpha ) (-4+\alpha ) (-2+\alpha ) \alpha  \left(1680-384 \alpha +260 \alpha ^2-44 \alpha ^3+5 \alpha ^4\right) \kappa _-^2\\
	&+15120 (-10+\alpha ) (-8+\alpha ) (-6+\alpha ) (-4+\alpha )
	(-2+\alpha ) \alpha  \left(20-2 \alpha +\alpha ^2\right) \kappa _-^4\\
	&-60480 (-10+\alpha ) (-8+\alpha ) (-6+\alpha ) (-4+\alpha ) (-2+\alpha ) \alpha  \kappa _-^6\\
	&-252 (-10+\alpha ) (-8+\alpha )
	(-6+\alpha ) (-4+\alpha ) (-2+\alpha ) \alpha  \left(1680-384 \alpha +260 \alpha ^2-44 \alpha ^3+5 \alpha ^4\right) \kappa _+^2\\
	&+1008 (-645120-4753664 \alpha +5538624 \alpha ^2-2371392 \alpha
	^3+607360 \alpha ^4\\
	&\qquad-116000 \alpha ^5+14968 \alpha ^6-1016 \alpha ^7+25 \alpha ^8) \kappa _-^2 \kappa _+^2\\
	&-20160 \left(-46080-96128 \alpha +112896 \alpha ^2-36088 \alpha ^3+4596 \alpha ^4-234
	\alpha ^5+3 \alpha ^6\right) \kappa _-^4 \kappa _+^2\\
	&-16128 (2+\alpha ) \left(8640-4952 \alpha +442 \alpha ^2+45 \alpha ^3\right) \kappa _-^6 \kappa _+^2\\
	&+15120 (-10+\alpha ) (-8+\alpha ) (-6+\alpha )
	(-4+\alpha ) (-2+\alpha ) \alpha  \left(20-2 \alpha +\alpha ^2\right) \kappa _+^4\\
	&-20160 \left(-46080-96128 \alpha +112896 \alpha ^2-36088 \alpha ^3+4596 \alpha ^4-234 \alpha ^5+3 \alpha ^6\right)
	\kappa _-^2 \kappa _+^4\\
	&-80640 \left(8640+952 \alpha -3264 \alpha ^2+500 \alpha ^3+15 \alpha ^4\right) \kappa _-^4 \kappa _+^4\\
	&+322560 \left(-720+530 \alpha -153 \alpha ^2+16 \alpha ^3\right) \kappa _-^6
	\kappa _+^4\\
	&-60480 (-10+\alpha ) (-8+\alpha ) (-6+\alpha ) (-4+\alpha ) (-2+\alpha ) \alpha  \kappa _+^6\\
	&-16128 (2+\alpha ) \left(8640-4952 \alpha +442 \alpha ^2+45 \alpha ^3\right) \kappa _-^2 \kappa
	_+^6\\
	&+322560 \left(-720+530 \alpha -153 \alpha ^2+16 \alpha ^3\right) \kappa _-^4 \kappa _+^6
	+258048 \left(-45+46 \alpha -18 \alpha ^2+2 \alpha ^3\right) \kappa _-^6 \kappa _+^6
	\Big)\\
\omega_{3, 3} &= 
-\frac{\irm\cot \left(\frac{\pi  \alpha }{2}\right)}{594542592(\alpha+1)(\alpha-1)(\alpha-3) \pi ^6}
\Big(
	(-6+\alpha ) (-4+\alpha ) (-2+\alpha ) \alpha  (2+\alpha ) (4+\alpha )\\
	&\qquad\times
	 \left(17856-2976 \alpha +1312 \alpha ^2+120 \alpha ^3+40 \alpha ^4-42 \alpha ^5+7 \alpha ^6\right)\\
	&-252 (-6+\alpha ) (-4+\alpha )(-2+\alpha ) \alpha  (2+\alpha ) (4+\alpha ) \left(336-48 \alpha +28 \alpha ^2-4 \alpha ^3+\alpha ^4\right) \kappa _-^2\\
	&+3024 (-6+\alpha ) (-4+\alpha ) (-2+\alpha ) \alpha  (2+\alpha ) (4+\alpha )
		\left(20-2 \alpha +\alpha ^2\right) \kappa _-^4\\
	&-12096 (-6+\alpha ) (-4+\alpha ) (-2+\alpha ) \alpha  (2+\alpha ) (4+\alpha ) \kappa _-^6\\
	&-252 (-6+\alpha ) (-4+\alpha ) (-2+\alpha ) \alpha  (2+\alpha )(4+\alpha ) \left(336-48 \alpha +28 \alpha ^2-4 \alpha ^3+\alpha ^4\right) \kappa _+^2\\
	&+1008 \left(-3072-148992 \alpha +47808 \alpha ^2+28352 \alpha ^3-9088 \alpha ^4+1088 \alpha ^5-88 \alpha ^6-40\alpha ^7+5 \alpha ^8\right) \kappa _-^2 \kappa _+^2\\
	&-12096 \left(-512-7808 \alpha +2592 \alpha ^2+1304 \alpha ^3-316 \alpha ^4-6 \alpha ^5+\alpha ^6\right) \kappa _-^4 \kappa _+^2\\
	&-48384 \left(64+288\alpha -132 \alpha ^2-12 \alpha ^3+3 \alpha ^4\right) \kappa _-^6 \kappa _+^2\\
	&+3024 (-6+\alpha ) (-4+\alpha ) (-2+\alpha ) \alpha  (2+\alpha ) (4+\alpha ) \left(20-2 \alpha +\alpha ^2\right) \kappa_+^4\\
	&-12096 \left(-512-7808 \alpha +2592 \alpha ^2+1304 \alpha ^3-316 \alpha ^4-6 \alpha ^5+\alpha ^6\right) \kappa _-^2 \kappa _+^4\\
	&-48384 \left(64+776 \alpha -368 \alpha ^2-20 \alpha ^3+5 \alpha^4\right) \kappa _-^4 \kappa _+^4-193536 \left(16-6 \alpha +3 \alpha ^2\right) \kappa _-^6 \kappa _+^4\\
	&-12096 (-6+\alpha ) (-4+\alpha ) (-2+\alpha ) \alpha  (2+\alpha ) (4+\alpha ) \kappa _+^6\\
	&-48384\left(64+288 \alpha -132 \alpha ^2-12 \alpha ^3+3 \alpha ^4\right) \kappa _-^2 \kappa _+^6-193536 \left(16-6 \alpha +3 \alpha ^2\right) \kappa _-^4 \kappa _+^6-774144 \kappa _-^6 \kappa _+^6
\Big)\\
\omega_{5, 1} &= 
	\frac{\irm\cot \left(\frac{\pi  \alpha }{2}\right)}{5945425920(\alpha-1)(\alpha+1)(\alpha+3) \pi ^6}
\Big(
	(-2+\alpha ) \alpha  (2+\alpha ) (4+\alpha ) (6+\alpha ) (8+\alpha )\\
	&\qquad\times \left(89280-15072 \alpha +4640 \alpha ^2-1224 \alpha ^3+608 \alpha ^4+294 \alpha ^5+35 \alpha ^6\right)\\
	&-252 (-2+\alpha ) \alpha 
	(2+\alpha ) (4+\alpha ) (6+\alpha ) (8+\alpha ) \left(1680-288 \alpha +116 \alpha ^2+4 \alpha ^3+5 \alpha ^4\right) \kappa _-^2\\
	&+15120 (-2+\alpha ) \alpha  (2+\alpha ) (4+\alpha ) (6+\alpha ) (8+\alpha
	) \left(20-2 \alpha +\alpha ^2\right) \kappa _-^4\\
	&-60480 (-2+\alpha ) \alpha  (2+\alpha ) (4+\alpha ) (6+\alpha ) (8+\alpha ) \kappa _-^6\\
	&-252 (-2+\alpha ) \alpha  (2+\alpha ) (4+\alpha ) (6+\alpha )
	(8+\alpha ) \left(1680-288 \alpha +116 \alpha ^2+4 \alpha ^3+5 \alpha ^4\right) \kappa _+^2\\
	&+1008 (-129024-1543936 \alpha -438720 \alpha ^2+281792 \alpha ^3+88960 \alpha ^4\\
	&\qquad+10528 \alpha ^5+3544
	\alpha ^6+616 \alpha ^7+25 \alpha ^8) \kappa _-^2 \kappa _+^2\\
	&-20160 \left(-9216-51328 \alpha -11328 \alpha ^2+8200 \alpha ^3+2436 \alpha ^4+198 \alpha ^5+3 \alpha ^6\right) \kappa _-^4 \kappa
	_+^2\\
	&-16128 (-4+\alpha ) \left(-864-2644 \alpha -712 \alpha ^2+45 \alpha ^3\right) \kappa _-^6 \kappa _+^2\\
	&+15120 (-2+\alpha ) \alpha  (2+\alpha ) (4+\alpha ) (6+\alpha ) (8+\alpha ) \left(20-2 \alpha
	+\alpha ^2\right) \kappa _+^4\\
	&-20160 \left(-9216-51328 \alpha -11328 \alpha ^2+8200 \alpha ^3+2436 \alpha ^4+198 \alpha ^5+3 \alpha ^6\right) \kappa _-^2 \kappa _+^4\\
	&-80640 \left(1728+5624 \alpha +96
	\alpha ^2-620 \alpha ^3+15 \alpha ^4\right) \kappa _-^4 \kappa _+^4\\
	&-322560 \left(144+110 \alpha +57 \alpha ^2+16 \alpha ^3\right) \kappa _-^6 \kappa _+^4\\
	&-60480 (-2+\alpha ) \alpha  (2+\alpha )
	(4+\alpha ) (6+\alpha ) (8+\alpha ) \kappa _+^6\\
	&-16128 (-4+\alpha ) \left(-864-2644 \alpha -712 \alpha ^2+45 \alpha ^3\right) \kappa _-^2 \kappa _+^6\\
	&-322560 \left(144+110 \alpha +57 \alpha ^2+16 \alpha
	^3\right) \kappa _-^4 \kappa _+^6-258048 \left(9-2 \alpha +6 \alpha ^2+2 \alpha ^3\right) \kappa _-^6 \kappa _+^6
\Big)
\end{align*}

\bibliographystyle{unsrt}
\bibliography{sample}

\end{document}